\documentclass{article}
\usepackage{graphicx}
\usepackage{color}      
\usepackage{dcolumn}    
\usepackage{bm}         
\usepackage[caption=false]{subfig}
\usepackage{float}
\usepackage{mathtools}
\usepackage{amsmath}
\usepackage{amssymb}
\usepackage{relsize}
\usepackage{geometry}
\usepackage{titlesec}   
\usepackage[sort&compress,numbers]{natbib}
\graphicspath{{./Figures/}}
\usepackage{soul}
\usepackage[dvipsnames]{xcolor}
\begin{document}


\begin{center}
    \textbf{\huge Active Convolved Illumination with Deep Transfer Learning for Complex Beam Transmission through Atmospheric Turbulence} \\[1em]
     Adrian A. Moazzam\textsuperscript{1}, Anindya Ghoshroy\textsuperscript{1}, Breeanne Heusdens\textsuperscript{2}, Durdu \"O.G\"uney\textsuperscript{1,3}, Roohollah Askari\textsuperscript{2} \\[0.5em]
     \textsuperscript{1}Department of Electrical and Computer Engineering, Michigan Technological University, Houghton, MI 49931, USA \\[0.5em]
     \textsuperscript{2}Geological and Mining Engineering and Sciences, Michigan Technological University, Houghton, MI 49931, USA \\[0.5em]
     \textsuperscript{3}John A. Paulson School of Engineering and Applied Sciences, Harvard University, Cambridge, MA 02138, USA \\[1em]
    \texttt{raskari@mtu.edu\\}
\end{center}


\begin{abstract}
Atmospheric turbulence imposes a fundamental limitation across a broad range of applications, including optical imaging, remote sensing, and free-space optical communication. Recent advances in adaptive optics, wavefront shaping, and machine learning, driven by synergistic progress in fundamental theories, optoelectronic hardware, and computational algorithms, have demonstrated substantial potential in mitigating turbulence-induced distortions. Recently, active convolved illumination (ACI) was proposed as a versatile and physics-driven technique for transmitting structured light beams with minimal distortion through highly challenging turbulent regimes. While distinct in its formulation, ACI shares conceptual similarities with other physics-driven distortion correction approaches and stands to benefit from complementary integration with data-driven deep learning (DL) models. Inspired by recent work coupling deep learning with traditional turbulence mitigation strategies, the present work investigates the feasibility of integrating ACI with neural network-based methods.
We outline a conceptual framework for coupling ACI with data-driven models and identify conditions under which learned representations can meaningfully support ACI’s correlation-injection mechanism. As a representative example, we employ a convolutional neural network (CNN) together with a transfer-learning approach to examine how a learned model may operate in tandem with ACI. This exploratory study demonstrates feasible implementation pathways and establishes an early foundation for assessing the potential of future ACI–DL hybrid architectures, representing a step toward evaluating broader synergistic interactions between ACI and modern DL models.

\end{abstract}

\section{Introduction}
Robust light transmission through spatiotemporally evolving disordered media, such as a turbulent atmosphere, remains fundamentally challenging due to the introduction of geometric distortions, phase aberrations, and spatial blurring of the original mode profile \cite{bertolotti2022imaging}. These effects impose significant limitations across a range of applications, including optical imaging, free-space communication, reconnaissance, and remote sensing. To address these bottlenecks, technologies such as adaptive optics \cite{hampson2021adaptive}, wavefront shaping, blind deconvolution \cite{levin2011understanding, campisi2017blind, harmeling2009online}, and machine learning \cite{lohani2018turbulence, feng2022turbugan} are increasingly being deployed. However, as operational demands expand into more extreme or variable propagation environments, the limitations of current systems are becoming more pronounced. This trend is fueling the rapid advancement of existing techniques and a renewed push for new or hybrid solutions \cite{schuler2013machine} that combine complementary approaches to achieve greater robustness and adaptability.

Recent advances in artificial intelligence (AI), particularly in deep learning, have opened new pathways for mitigating turbulence-induced distortions. AI-driven frameworks, often based on deep neural networks, are now being actively explored as augmentative tools to enhance the performance of conventional correction schemes. In adaptive optics systems, for instance, accurate wavefront sensing is critical for reliable phase reconstruction. Yet traditional approaches such as Shack-Hartmann wavefront sensors (SHWFS), shear interferometers, and pyramid wavefront sensors (PWFS) are constrained by trade-offs between sensitivity and precision. Furthermore, once these systems are fabricated, their performance characteristics are typically fixed, limiting their ability to respond to dynamically changing environmental conditions. By incorporating deep learning models into wavefront sensing and control pipelines, it becomes possible to not only improve correction fidelity but also enable real-time adaptability to evolving turbulence regimes. A growing body of work has begun to explore and validate such hybrid systems, as surveyed in \cite{liu2025research,hill2025deep}.

Machine learning has swiftly emerged as a ubiquitous and pervasive strategy for tackling complex challenges across diverse domains. Researchers have developed numerous methods to better understand, simulate, and mitigate its detrimental effects on imaging systems. Foundational efforts favored physically grounded simulations: for instance, Repasi and Weiss \cite{repasi2011computer} devised a method that incorporates range-dependent turbulence for horizontal views, whereas Schwartzman et al. \cite{schwartzman2017turbulence} introduced a 2D correlated distortion model to circumvent resource-intensive 3D computations. More recent work leverages machine learning to bolster both realism and computational speed. Miller et al. \cite{miller2019data} pioneered a data-constrained random walk algorithm for real-time blur and distortion, while Miller and Du Bosq \cite{miller2021machine} extended this framework by employing generative adversarial networks (GANs) to more accurately capture spatially varying turbulence effects. 

Substantial progress has also been achieved on the restoration front, where DL methods are increasingly employed to correct turbulence-induced distortions. One approach employs two-stage adversarial networks to tackle geometric distortion and blur individually \cite{rai2022removing}, augmenting restoration efficacy with sub-pixel and channel-attention modules. Other researchers have applied residual learning-based convolutional neural networks (CNNs) to refine de-blurring and contrast in video sequences \cite{gao2019atmospheric}, while CNNs have similarly been leveraged to remedy optical beam distortions and boost fidelity in orbital angular momentum signals \cite{liu2019deep, zhai2020turbulence}. Guo et al. leverage deep transfer learning to facilitate blind restoration \cite{guo2022blind}. Further developments involve transformer-based architectures that decouple turbulence physics from the restoration process and employ temporal attention for improved spatio-temporal modeling \cite{zhang2024imaging}, as well as physics-driven approaches that integrate simulations and stochastic refinements to ensure robust performance under real-world conditions \cite{jaiswal2023physics}. Moreover, recent work such as Restormer \cite{zamir2022restormer} has demonstrated the potential of efficient transformer frameworks for high-resolution image restoration. Another promising avenue employs U-Net-like deep autoencoders to address both geometric distortion and blur in severely degraded images \cite{chen2019u}.

Active convolved illumination (ACI) is an optical technique that has so far shown to compensate for intrinsic system limits, distortion, noise, attenuation, and absorption, across near-field coherent/incoherent and far-field incoherent regimes \cite{sadatgol2015plasmon,ghoshroy2020loss,adams2016review,adams2016bringing,zhang2016enhancing,zhang2017analytical,adams2017plasmonic,adams2018plasmonic,adams2022incoherent,ghoshroy2020theory}. More recently, ACI was introduced as a versatile technique for transmitting structured light beams with minimal distortion through turbulent media \cite{ghoshroy2024enhancing, moazzam2025remote, moazzam2023active}. This implementation probes distinct regions of the diffraction cone using a basis of orbital-angular-momentum (OAM) modes to sample the distortions. A reciprocal-space analysis then identifies correlation-injecting sources (CISs), narrow-band beams that, when convolved with the object, form an auxiliary source that counteracts these distortions, yielding high-fidelity transmission and reconstruction even under moderate-to-severe turbulence. This technique maintains the geometric integrity of optical beams with complex mode profiles during propagation, which is particularly challenging, as most structured light modes are highly susceptible to turbulence-induced aberrations. Despite this vulnerability, such modes remain highly desirable for applications in high-resolution imaging, remote sensing, and optical communications. For instance, orthogonal mode sets such as Laguerre-Gaussian, Bessel, Hermite-Gaussian, and Ince-Gaussian modes, as well as their vectorial combinations, have been widely proposed for high-capacity free-space communication links. However, these modes face a critical limitation. Their geometric structure degrades severely during long-distance transmission through volumetric turbulence, resulting in loss of orthogonality. Consequently, practical implementations of both classical and quantum communication channels using these modes are typically restricted to relatively short ranges \cite{singh2023robust}. In contrast, ACI offers spatial mode multiplexing opportunities where deployment of orthogonal mode sets, turbulence resilient, self-healing, or propagation invariant eigenmodes may be challenging.

Although the proposed ACI framework has demonstrated resilience across a broad dynamic range of propagation conditions, it remains susceptible to limitations analogous to those encountered in adaptive optics and wavefront shaping systems. While distinct in its formulation, ACI shares conceptual parallels with adaptive optics in its goal of mitigating turbulence-induced degradation by actively modifying the input field. However, unlike adaptive optics systems that rely on closed-loop feedback and direct wavefront measurements, ACI operates through a forward-designed correlation injection strategy that encodes resilience into the beam itself via a convolution with a carefully synthesized correlation injecting source (CIS). A sequence of CISs is constructed by propagating a series of orbital angular momentum (OAM) modes with varying reciprocal space parameters. This process can be interpreted as segmenting an arbitrarily broad diffraction cone, and each OAM mode effectively samples the distortion space. A spatiotemporal snapshot of the conjugate of the turbulence-induced distortion is then encoded into each CIS. Through convolution, the encoded CIS is injected into beams with arbitrary mode profiles that traverse through the sampled diffraction cone. However, due to the chaotic nature of atmospheric turbulence, even minor errors in the CIS synthesis can lead to significant performance degradation, particularly under anisoplanatic conditions. This sensitivity and the pilot-beam-like nature of the CISs, suggest a conceptual overlap with techniques used in adaptive optics and wavefront shaping, and opens avenues for complementary integration with data-driven deep learning models \cite{liu2024atmospheric}.


Inspired by recent advances in integrating deep learning with traditional turbulence-mitigation strategies, this work investigates how data-driven models can be incorporated into the ACI framework. Our aim is to outline a potential pathway for coupling ACI with deep learning, characterize the information flow such a system may require, and identify conditions under which learned models could complement the ACI correlation-injection process.
To illustrate this conceptual pathway, we implement a CNN-based model alongside ACI as a representative example. This case study allows us to explore how learned representations may interact with the ACI workflow, what capabilities a DL model must possess to function coherently within correlation-injection strategies. Several aspects of this exploration are intentionally forward-looking, reflecting speculative but physically motivated considerations for future ACI–DL hybrid systems.
We further present results from a CNN initially trained on diverse datasets containing non-atmospheric noise characteristics and subsequently fine-tuned for atmospheric imaging tasks, demonstrating one practical route through which deep learning could be incorporated into ACI-based imaging pipelines.

In the Simulation Framework and Dataset section, we detail the numerical simulations utilized to represent atmospheric distortion and describe the architecture of our CNN model and the utilized datasets. The Results and Discussion section provides a comprehensive comparative analysis of performance, examining the efficacy of ACI alone, the DL model alone, and our example hybrid implementation. This section also considers the difficulties and challenges that deep learning must overcome to achieve improved performance in atmospheric imaging. Finally, the Conclusion summarizes key findings, emphasizing that this work outlines a pathway for potential ACI–DL integration, and outlines potential directions for future improvements.

\section{Turbulence Distortion Characterization}
Turbulence arises in fluid flow when inertial forces dominate over viscous forces \cite{warhaft2002turbulence}. In Earth's atmosphere, this phenomenon significantly affects the propagation of electromagnetic waves, causing considerable challenges for optical imaging systems. Atmospheric turbulence predominantly results from temperature gradients (thermal convection) and wind shear, creating random fluctuations in the refractive index of air \cite{andrews2005laser, roggemann2018imaging}. These refractive index variations distort the wavefront of light traveling through the atmosphere, introducing multiple detrimental effects on optical systems.

In optical imaging, atmospheric turbulence induces several critical distortions. Beam wandering, characterized by random off-axis shifts of the beam centroid due to turbulent eddies \cite{andrews2005laser}. Additionally, wavefront aberrations manifest in both lower-order forms, such as tilt and defocus, and higher-order complexities, significantly impairing image quality \cite{roggemann2018imaging, hardy1998adaptive}. Turbulence also causes geometric deformations that irregularly distort and stretch images, altering the perceived shapes of observed objects \cite{fried1966optical}. Moreover, scintillation—rapid intensity fluctuations resulting from small-scale refractive index irregularities—further complicates accurate imaging \cite{andrews2005laser, tatarskii1971effects}. Lastly, random phase fluctuations lead to image blurring, reducing both spatial resolution and contrast \cite{fried1966optical, roddier1981v}.

Collectively, these atmospheric turbulence effects pose substantial limitations to ground-based imaging systems, necessitating the development of advanced mitigation methods. Optical strategies such as adaptive optics \cite{hampson2021adaptive,wang2018performance}, structured light \cite{cox2020structured}, and ACI \cite{ghoshroy2024enhancing,moazzam2025remote} have effectively improved image clarity and resolution. Additionally, computational post-processing approaches, including Lucky imaging \cite{fried1978probability,zhang2011astronomical}, and iterative image reconstruction methods \cite{furhad2016restoring,zhu2010image}, offer complementary ways to enhance imaging outcomes. Recent advancements in machine learning and deep learning further expand these capabilities, demonstrating remarkable potential in addressing turbulence-induced image degradation. Notably, hybrid approaches are also emerging: for instance, Lv et al. \cite{lv2025adaptive} combined Lucky imaging with GANs to leverage both temporal frame selection and generative restoration for enhanced performance.

In the following sections, we detail our novel integration of ACI with the CNN-based deep learning model, highlighting the significant improvements achievable in atmospheric imaging quality through this synergistic approach.

\section{Simulations Framework and Dataset}

\subsection{Atmospheric Distortion Dataset Simulation Framework}
\label{Atmospheric_Simulation}

To generate a synthetic dataset that models atmospheric distortions and illustrates the impact of turbulence on random coherent objects, we simulated their propagation through a turbulent medium using the split-step method \cite{schmidt2010numerical}, along with von Kármán's power spectral density model. In our framework, the outer scale $L_0 = 100 \textrm{m}$ and the inner scale $l_0 = 1 \textrm{cm}$ statistically represent the typical sizes of the largest and smallest turbulent eddies, respectively \cite{ke2023generation}. The objects were propagated over a distance of $L = 1 \textrm{km}$. With the propagation distance fixed and using a monochromatic wavelength of $\lambda = 1550\textrm{nm}$, we investigated various turbulence regimes by adjusting the refractive index structure parameter $C_n ^2$ to values of $0.7 \times 10^{-14}$, $1.7 \times 10^{-14}$, and $2.7 \times 10^{-14}\textrm{m}^{-2/3}$.

In our simulations, we used a grid of 8192 × 8192 pixels, where each pixel represents a physical size of 0.35 mm on both the source and observation planes. This level of spatial sampling was selected to achieve high accuracy with the angular spectrum propagation technique under turbulent conditions. Additionally, to generate input for our DL model, we cropped the observation window to a central 512 × 512 region. This strategy ensured that all key features of the target objects were preserved while reducing the dataset size to a level that allows for efficient computational processing.

Turbulence severity is typically quantified using the conventional $D/r_0$ metric, where $D$ denotes the imaging aperture size and $r_0$ is the Fried parameter. In our study, however, we employed an alternative measure, $\Delta l_t/\Delta l_s$ \cite{ghoshroy2024enhancing}. In this formulation, the atmospheric resolution limit is defined as $\Delta l_t = 1.22\,L\lambda/r_0$, while $\Delta l_s = 1.22\,L\lambda/D$ represents the resolution imposed by the aperture size. Here, $D$ is taken as the effective diameter of a phantom aperture that collects approximately $1-1/e^2$ of the total diffraction power at the observation plane. Furthermore, we evaluated anisoplanatism by calculating the ratio $\theta_s/\theta_t$, where $\theta_s$ represents the object size at the source plane and $\theta_t$, (determined as the product of the isoplanatic angle and the propagation distance $L$) defines the isoplanatic patch.

To evaluate the effects of turbulence and the efficacy of our correction methods, we employed the (NCC) metric, $\xi$, as an indicator of distortion. Specifically, we denote the NCC value for wave propagation without any correction as $\xi_P$. When ACI is applied, the corresponding metric is labeled as $\xi_A$. Additionally, when distortion is corrected using our DL model, we refer to the resulting NCC value as $\xi_{DL}$. Finally, if both ACI and the DL model are applied, the combined correction is represented by $\xi_{A+DL}$.

\subsection{Data Set}

Training neural networks effectively demands careful fine-tuning of model weights, a process highly dependent on the quality and relevance of the training dataset. In this study, considering that our network utilized transfer learning with pre-trained weights, an excessively large dataset was unnecessary \cite{yosinski2014transferable}. Instead, our priority was to generate a sufficiently diverse yet representative set of target patterns.

To achieve this, we constructed random target objects characterized by multiple degrees of freedom. Each parameter defining these objects was methodically varied within carefully chosen ranges, ensuring the generated targets neither appeared overly large, thus reducing the sensitivity to atmospheric turbulence effects, nor excessively small, risking violation of sampling requirements. This deliberate approach to random target generation enabled us to produce a dataset that robustly captures a broad spectrum of potential imaging scenarios \cite{torralba2011unbiased, sun2017revisiting}.

\begin{figure*}[th]
\centering
\includegraphics[width=0.85\linewidth]{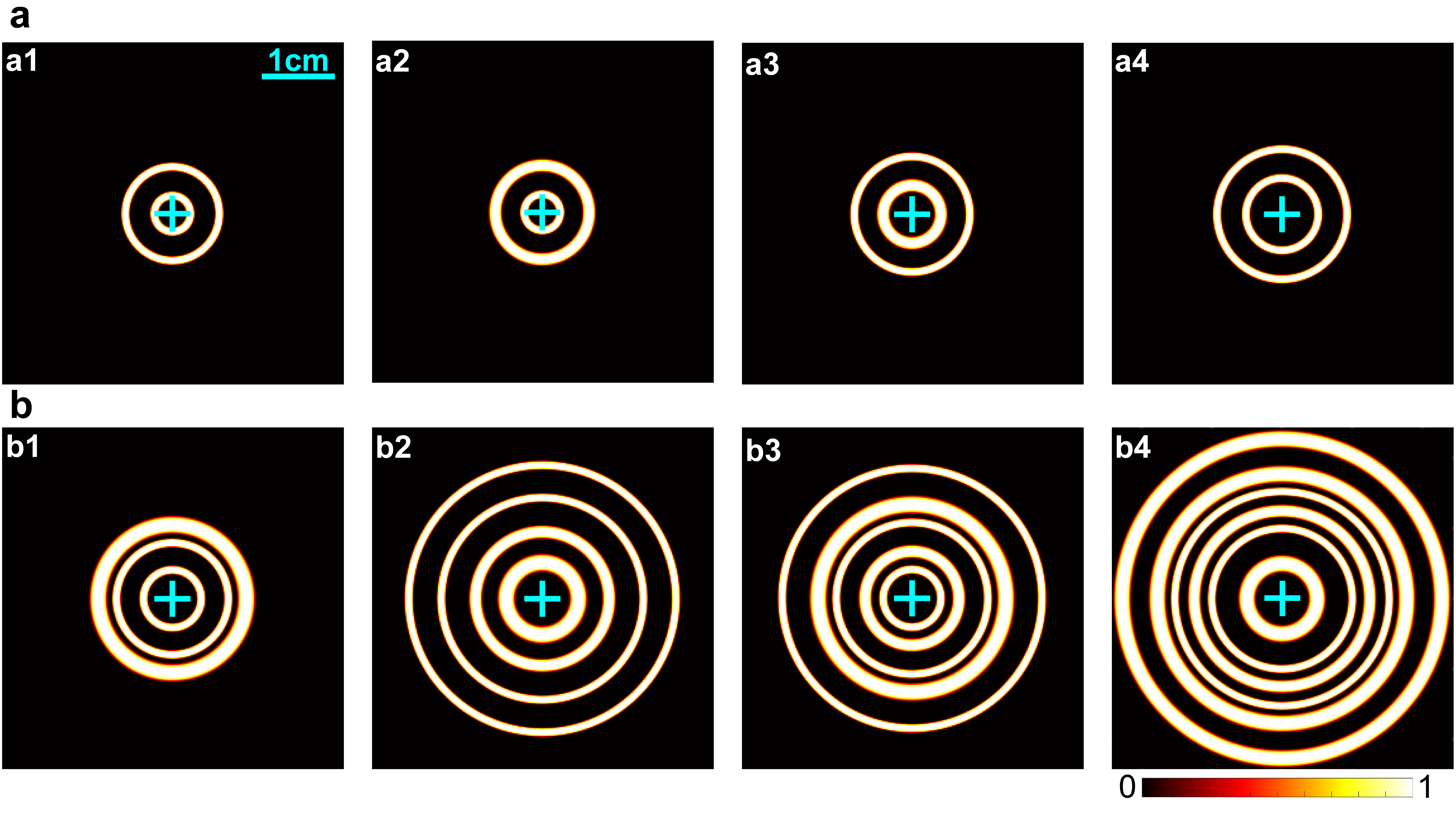}
\caption{The dataset used for training the DL model was synthetically generated using randomly generated concentric ring patterns. For each target, the diameter and radial position of each ring were independently sampled within predefined degrees of freedom, resulting in variations in ring width and spacing, even among targets with the same number of rings, as illustrated in row (a). Additionally, the total number of rings per target was randomly selected, further increasing structural diversity, as shown in row (b). This design enables the model to generalize across a wide range of spatial frequencies and geometric configurations.}
\label{fig:Data_set}
\end{figure*}

We generated 100 unique random targets similar to the examples depicted in Fig. \ref{fig:Data_set}. These targets comprised rings with varying widths and radii, as illustrated in Fig. \ref{fig:Data_set}(a), and also included differing numbers of rings, as shown in Fig. \ref{fig:Data_set}(b). Each generated target was subsequently propagated through a turbulent atmosphere five separate times, each propagation utilizing a fresh and distinct realization of turbulence. After propagation, each instance was reconstructed, resulting in a passive dataset totaling 500 unique examples.

In parallel, the same set of 100 random targets underwent propagation through the same atmospheric turbulence model, but this time incorporating ACI to actively enhance target fidelity. These propagations, also reconstructed, yielded an active dataset comprising another 500 examples. This systematic and balanced data acquisition strategy ensured equitable representation and robust statistical significance across varying atmospheric conditions \cite{he2009learning, buda2018systematic}, thereby enabling a comparative analysis of the integrated ACI and DL framework. In both passive and active cases, we included an identical set of three different $C_n^2$ values. This means 6 unique datasets in all. Each dataset was split into training and test sets using a 9:1 ratio, and the model was trained/tested separately on each. In cases where both passive and active datasets were subjected to the same $C_n^2$ value, the reconstructed passive targets exhibited higher levels of distortion, whereas the active targets showed lower levels of distortion. Consistent with prior reports \cite{nieuwenhuizen2019deep}, we therefore trained the model separately on each dataset, with our DnCNN achieving slightly better performance under this approach.

 \subsection{Architecture and Training of the Deep Transfer Learning Model}

Our denoising framework is based on the Denoising Convolutional Neural Network (DnCNN) architecture \cite{zhang2017beyond}, known for its effectiveness in learning residual mappings for noise removal. The network begins with a 2D convolutional layer using 64 filters of size $3\times3$, followed by a ReLU activation. This is succeeded by 18 repeated blocks of convolution, batch normalization, and ReLU layers, all using $3\times3$ filters with stride 1 and padding to preserve spatial resolution. A final convolutional layer estimates the noise, which is subtracted from the input to produce the denoised output.

To adapt DnCNN for atmospheric imaging, we adopted a transfer learning strategy with two main architectural modifications: (1) adjusting the input layer to match our data dimensions, and (2) modifying the output layer to improve generalization. In general, the performance of base models can be improved through transfer learning, including for atmospheric applications with limited datasets \cite{10062031}.  DnCNN has been shown to perform well when fine-tuned for more specific denoising tasks without introducing new artifacts \cite{UsuiKeisuke2021Qeod}.
The DnCNN model was trained in two stages. First, it was pretrained on 48,000 grayscale images with synthetic Gaussian noise to learn general noise features. Then, it was fine-tuned on a smaller dataset featuring atmospheric distortions, allowing the network to adapt to the specific characteristics of turbulent image degradation.

During training, we used mini-batch gradient descent with a batch size of 32, along with early stopping based on validation loss calculated using the structurally emphasized SSIM (SE-SSIM) metric. A learning rate scheduler reduced the learning rate from $10^{-2}$ to $10^{-6}$ to improve convergence. We kept the model weights unfrozen during fine-tuning to allow full parameter optimization, improving performance over fixed-weight approaches.

\begin{figure*}[th]
\centering
\includegraphics[width=0.45\linewidth]{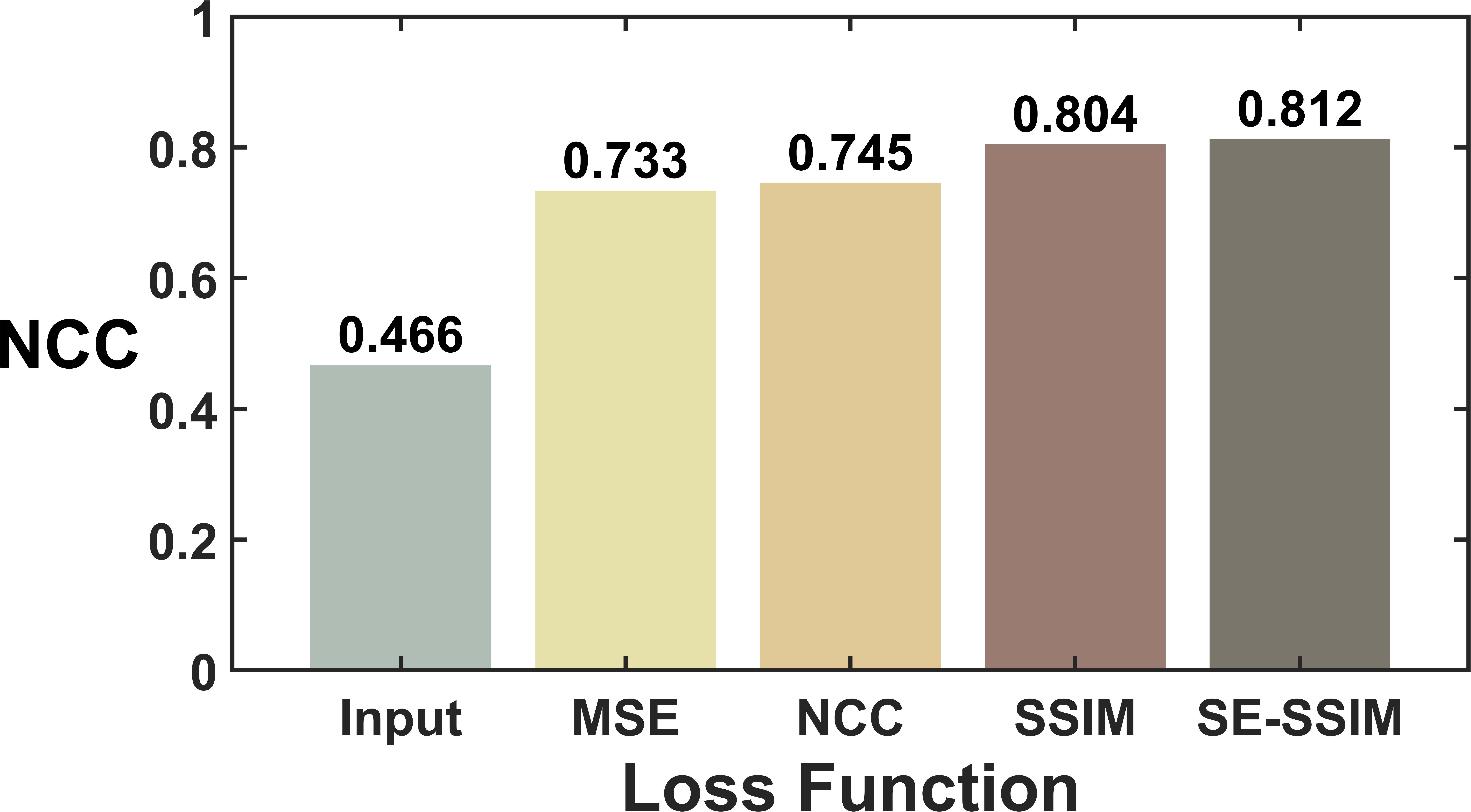}
\caption{The average NCC between reconstructed outputs and their corresponding ground truth targets across the test set is shown. The "Input" refers to the test set without any DL enhancement (for $C_n^2 = 0.7 \times 10^{-14}\,\textrm{m}^{-2/3}$), where NCC is computed for each target and averaged across the set. Multiple loss functions were evaluated for training the DL model, and the performance of each configuration was assessed by comparing the DL outputs to the ground truth using NCC. The resulting average NCC for each loss function is displayed for comparison.}
\label{fig:Loss_function}
\end{figure*}

Loss function selection played a critical role in model effectiveness, recognizing that the selection of a loss function is vital for effective image restoration, even within a single network \cite{zhao2016loss}. Beyond traditional mean squared error (MSE) \cite{zhang2017beyond, snell2017learning}, we explored normalized cross-correlation (NCC) and SSIM-based loss functions, including standard SSIM and SE-SSIM. The SSIM between two image patches, $I_n$ (noisy image) and $I_t$ (ground truth image), is defined as $SSIM(I_n,I_t) = [l(I_n,I_t)]^\alpha .[c(I_n,I_t]^\beta . [s(I_n,I_t)]^\gamma$, where $l$, $c$, and $s$ represent the luminance, contrast, and structural similarity components, respectively \cite{wang2004image}.  In SE-SSIM, the exponents are modified to emphasize structural similarity, with $\alpha = 0.1$, $\beta = 0.1$, $\gamma = 0.8$, making it more sensitive to perceptual distortions caused by atmospheric turbulence. As shown in Fig. \ref{fig:Loss_function}, different loss formulations yield varying improvements in normalized cross-correlation (NCC) with the ground truth. As illustrated in Fig. \ref{fig:Loss_function}, a comparative analysis of the model’s performance using these different loss formulations highlights the improvements in image quality achieved by each method. We should note that the same model parameters were used for all 6 datasets.

The DnCNN architecture \cite{zhang2017beyond} adopted in this work is a well-established and widely studied model, originally developed for image restoration tasks such as denoising, super-resolution, and deblurring to replace traditional blur estimation and unsharp masking operations. It has since been adapted for turbulence mitigation in optical imaging \cite{nieuwenhuizen2019deep,gao2019atmospheric}, making it a simple yet relevant albeit relatively dated choice for the present application. Although DnCNN is known to underperform in tasks involving complex geometric distortions such as warping and structural deformation, it remains a commonly used baseline in the literature and serves as a reference point for evaluating more advanced architectures \cite{vint2020analysis}. In this work, DnCNN is employed not as an optimal solution, but as a controlled benchmark to explore the compatibility between deep learning models and the ACI framework.

The transfer learning strategy here is selected to introduce some overlap with the DnCNN architecture with the operating regime for ACI. This is achieved by bridging the statistical structure of DnCNN's training priors with turbulence-induced distortions observable under the propagation conditions of ACI. This enables meaningful analysis in the present work without overfitting to a specific model.

The rationale for incorporating a transfer learning strategy is grounded in the statistical differences between typical image restoration datasets used in DL models and the physical distortions introduced by propagation through turbulence. DnCNN, having been pretrained on large-scale natural image datasets for tasks like denoising and deblurring, has already internalized a set of generic features, such as edge enhancement, and low-level spatial geometric correlations, that are still partially relevant to the types of distortions observed in optical imaging through turbulence. However, the distortion statistics under which ACI operates are not fully aligned with these priors. For example, ACI is intended to operate under larger $\Delta l_t/\Delta l_s$ where higher-order aberration effects and severe geometric distortion are dominant. Rather than training a model from scratch, which would require large, domain-specific datasets and risk overfitting, transfer learning allows us to adapt an existing DnCNN architecture to a new domain while preserving its generalization capability. By fine-tuning DnCNN on the typical operating conditions for ACI, we exploit the overlap between the model's learned representations and the propagation-induced variations for which ACI is designed. This enables the network to specialize its responses without requiring a completely new architecture, and offers insights into how conventional CNNs can be made to interact with the ACI framework. Thus, the use of transfer learning is not merely a workaround but a deliberate strategy to investigate the compatibility and potential synergy between existing data-driven DL models and physics-based distortion mitigation frameworks like ACI.

While the present work adopts standard reference metrics such as NCC or SSIM to assess reconstruction quality against known ground truth targets, no-reference image quality metrics such as  Natural Image Quality Evaluator (NIQE) \cite{mittal2012making}, and the Blind Image Spatial Quality Evaluator (BRISQUE) \cite{mittal2012no} may be better alternatives to reflect perceived image fidelity under more realistic, unsupervised conditions.

\section{Results and Discussion}

In this section, we investigate two distinct implementations integrating ACI with our DL model, focusing on their complementary capabilities and potential avenues for future development. Additionally, we discuss the other potential applications of DL models in conjunction with ACI, aiming to further enhance the synergy between optical and post-processing approaches.

\subsection{DnCNN Integration with ACI}

To ensure meaningful comparisons, randomly generated targets in our dataset were analyzed under conditions of matched isoplanatic patches ($\theta_s/\theta_t$) and a fixed refractive index structure constant ($C_n^2 = 0.7 \times 10^{-14}, \textrm{m}^{-2/3}$), while the turbulence severity ($\Delta l_t/\Delta l_s$) varies as these targets are illustrated in Fig. \ref{fig:Turb_Severity_level}. The ground truth of the targets are presented in Fig. \ref{fig:Turb_Severity_level}, column a. As the turbulence severity increases from top to bottom rows, the passive reconstruction method (Fig. \ref{fig:Turb_Severity_level}, column b) demonstrates decreasing NCC ($\xi_P$) values. However, these outcomes remain sensitive to individual short-exposure atmospheric realizations. Integrating the DL model alone (Fig. \ref{fig:Turb_Severity_level}, column c) enhances robustness by mitigating distortions caused by atmospheric turbulence, as indicated by higher NCC values ($\xi_{DL}$) compared to passive scenarios, as it consistently outperforms passive reconstruction. Though the performance of the DL model slightly decreases with increasing turbulence severity. ACI alone (Fig. \ref{fig:Turb_Severity_level}, column d) preserves the structural integrity of targets, reducing the impact of increasing turbulence severity, as reflected by NCC values ($\xi_{A}$). The combined use of ACI and the DL model (Fig. \ref{fig:Turb_Severity_level}, column e) further elevates image fidelity, consistently yielding higher NCC values ($\xi_{A+DL}$). However, it is important to note that although NCC is used as a comparative metric and may indicate improvements, its values should not be interpreted literally as direct enhancement in image quality due to the reconstruction artifacts arising from the DnCNN inadequacies, especially in highly distorted cases where intensity-based metrics can mask genuine structural recovery.
Nevertheless, this synergy highlights the complementary strengths of optical and post-processing methods, where ACI provides initial structural preservation and the DL delivers fine-level refinement.

\begin{figure}[H]
\centering
\includegraphics[width=0.89\linewidth]{Figures/Targets_cn2_7e15_v3.png}
\caption{Three randomly generated targets composed of the same number of concentric rings but with varying feature sizes are presented. Each target was selected to correspond to similar isoplanatic patch ($\theta_s/\theta_t$) conditions while leading to different turbulence severity ($\Delta l_t/\Delta l_s$) during atmospheric propagation. Column (a) shows the ground truth targets. Column (b) presents the targets after propagation through atmospheric turbulence with a refractive index structure constant of $C_n^2 = 0.7 \times 10^{-14}\,\textrm{m}^{-2/3}$. Column (c) shows the outputs from our DL model applied to the results in column (b). In column (d), the targets are reconstructed after propagation through the same atmospheric conditions with ACI applied. The reconstructions in column (d) are then processed by the DL model, yielding the final results in column (e). The NCC between each reconstruction and its ground truth is denoted by $\xi$.}
\label{fig:Turb_Severity_level}
\end{figure}

\begin{figure}[H]
\centering
\includegraphics[width=0.89\linewidth]{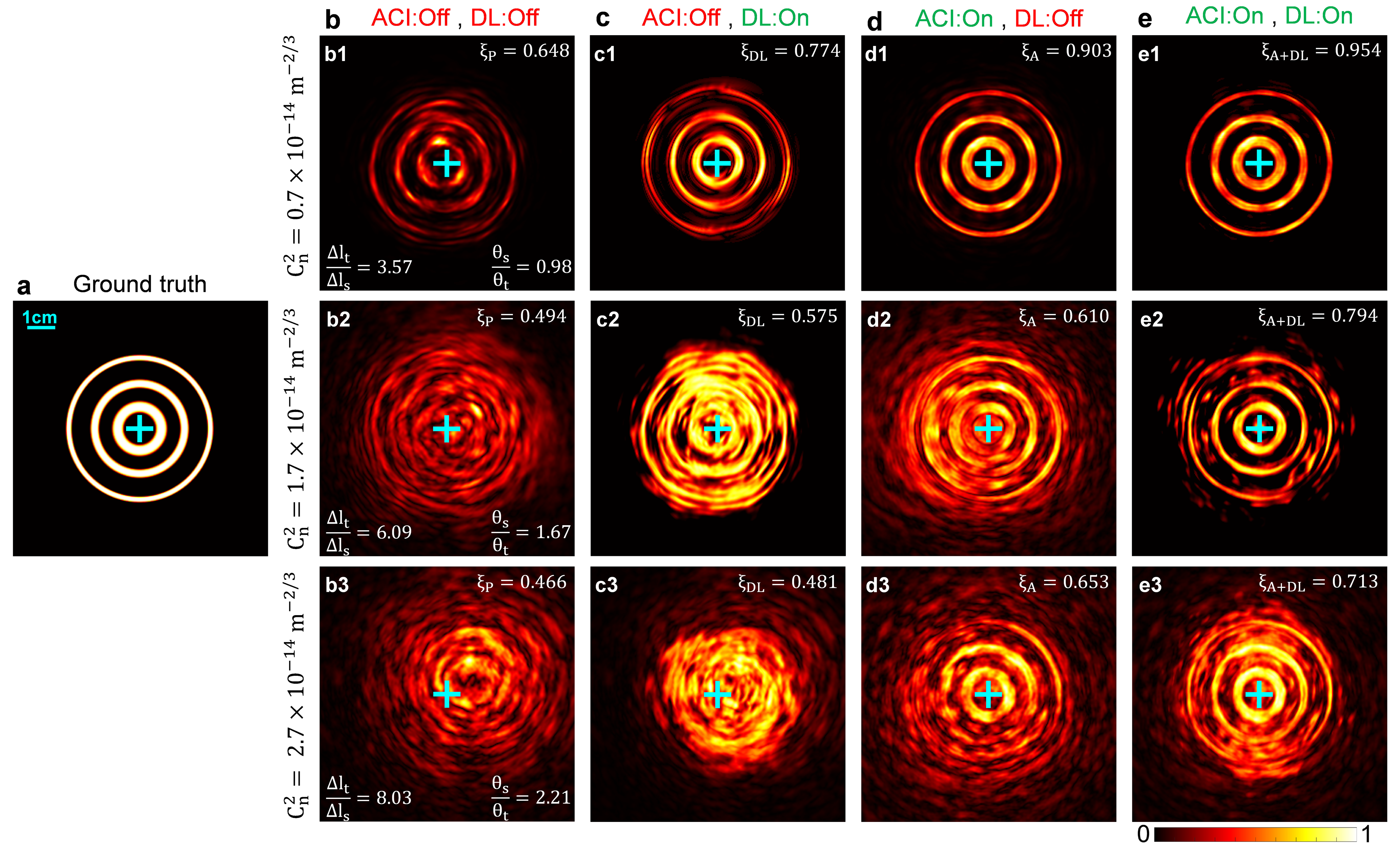}
\caption{A representative target is propagated through atmospheric turbulence and reconstructed under various scenarios for three different turbulence levels ($C_n^2 = 0.7 \times 10^{-14}$, $1.7 \times 10^{-14}$, and $2.7 \times 10^{-14}\,\textrm{m}^{-2/3}$). Column (a) shows the ground truth target. Column (b) presents the target after propagation through turbulence. Column (c) displays the output of our DL model applied to the results in column (b). Column (d) shows reconstructions with ACI applied during propagation. These ACI-enhanced reconstructions are further refined using the DL model, with the final outputs shown in column (e). In all columns, turbulence distortion severity increases from rows 1 to 3. The NCC between each reconstruction and the ground truth is denoted by $\xi$.}
\label{fig:Turb_all_cn2_level}
\end{figure}

Since targets recovered using ACI show considerable improvement, we further extend our analysis by evaluating system performance and investigate the impact on target integrity across progressively higher turbulence conditions ($C_n^2 = 0.7 \times 10^{-14}$, $1.7 \times 10^{-14}$, and $2.7 \times 10^{-14} \textrm{m}^{-2/3}$). The performance of models employing only DL, only ACI, and the combined approach (ACI with DL) across varying turbulence conditions is presented in Fig. \ref{fig:Turb_all_cn2_level}. Ground truth references are depicted in Fig. \ref{fig:Turb_all_cn2_level}(a), followed by passive cases (b), passive combined with DL cases (c), active cases (d), and active combined with DL cases (e). Results demonstrate that as atmospheric turbulence intensifies, the advantages of using ACI alone and integrating the DL model with ACI become increasingly pronounced.

Based on our observations, the DL model primarily mitigates ripple effects, enhancing the visibility and clarity of structures when they remain at least partially detectable, as seen in Fig. \ref{fig:Turb_Severity_level}(c1–c3) and Fig. \ref{fig:Turb_all_cn2_level}(c1). At higher distortion levels, however, where structural integrity is severely compromised, the model fails to effectively reconstruct the target (Fig. \ref{fig:Turb_all_cn2_level}(c2–c3)). In certain cases, distortions adjacent to structures are inadvertently accentuated, as evident in Fig. \ref{fig:Turb_all_cn2_level}(c2–c3). A similar effect is observed in the ACI with DL case (bottom of Fig. \ref{fig:Turb_all_cn2_level}(e3)), where distortions appear as part of the target itself. This lack of geometric awareness in both passive and active results may stem from suboptimal training with complex geometric distortions, compounded by DnCNN’s known limitations in modeling such effects. Higher-capacity architectures, including U-Net \cite{ronneberger2015u}, SwinIR \cite{liang2021swinir}, EDSR \cite{lim2017enhanced}, Restormer \cite{zamir2022restormer} and transformer-based models, could better capture these distortions and enhance turbulence mitigation.

\begin{figure}[H]
\centering
\includegraphics[width=0.98\linewidth]{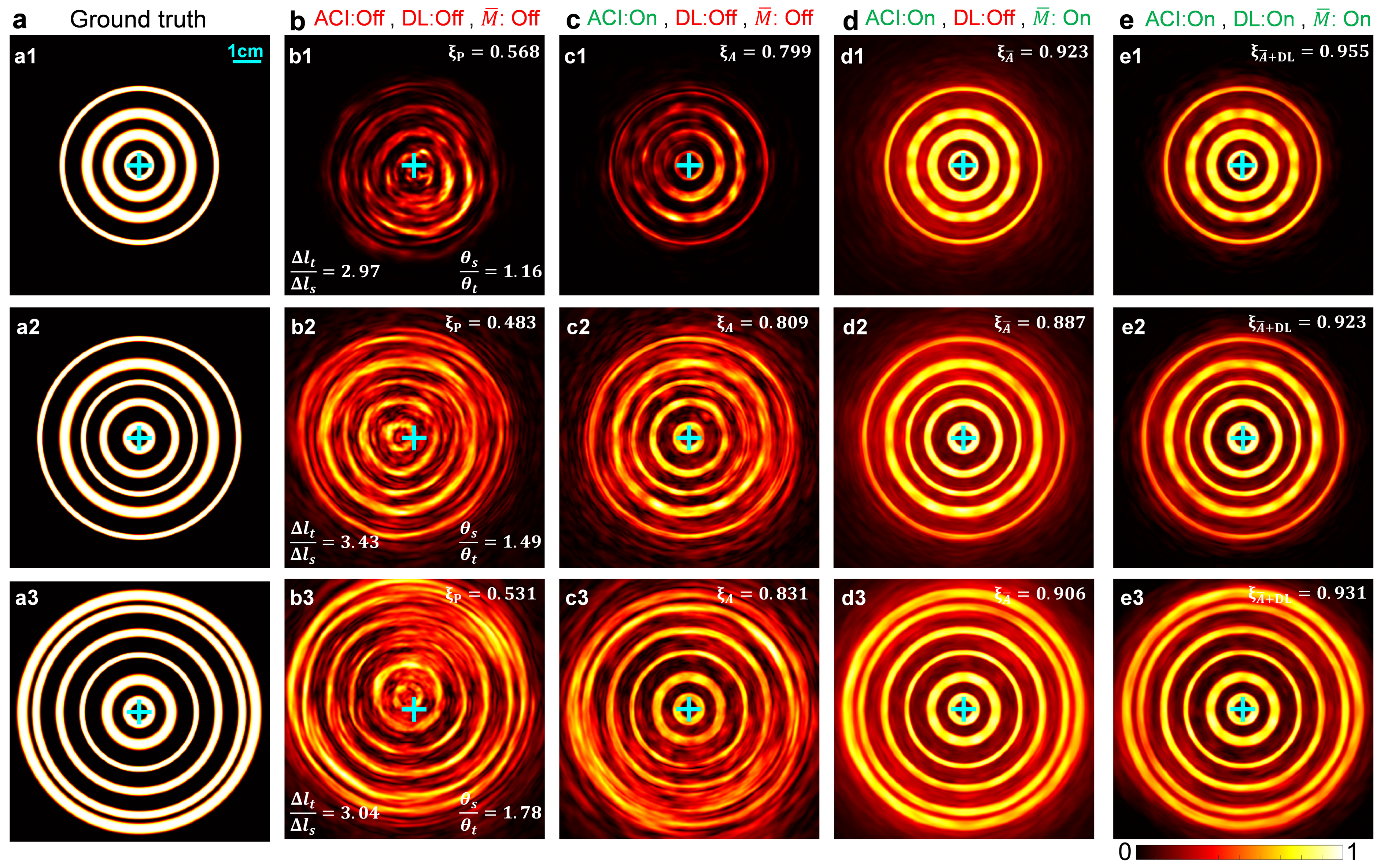}
\caption{Three targets with various numbers of rings are shown, each propagated under a fixed turbulence strength of $C_n^2 = 0.7 \times 10^{-14}\,\textrm{m}^{-2/3}$. Column (a) shows the ground truth targets. Column (b) presents the targets after propagation through atmospheric turbulence and subsequent reconstruction. In column (c), the targets are reconstructed with ACI applied during propagation. Column (d) shows the ACI-based reconstructions averaged over ten independent atmospheric realizations. These averaged results are further refined using the DL model, with the final outputs shown in column (e). The NCC between each reconstruction and its corresponding ground truth is denoted by $\xi$.}
\label{fig:Average_ACI_DL_Cn2_7e-15}
\end{figure}

In passive cases, our DL model also reduces beam wandering (random off-axis centroid shifts) similar to the active cases (with or without DL), particularly under moderate turbulence (Fig. \ref{fig:Turb_Severity_level}(c1–c3)), though residual wandering remains visible at stronger turbulence levels (Fig. \ref{fig:Turb_all_cn2_level}(c3)), consistent with prior findings by Gao et al. \cite{gao2019atmospheric}. The performance diminishes due to dominant higher- and lower-order aberrations as well as other geometric deformations (Fig. \ref{fig:Turb_all_cn2_level}(c2–c3)). While ACI alone can partially correct these distortions, it also struggles under severe turbulence effect (high $\Delta l_t/\Delta l_s$ values) despite the DL integration (Fig. \ref{fig:Turb_all_cn2_level}(e3)). Consequently, our DL model similarly shows limited capability in correcting severe aberrations and large-scale geometric distortions, particularly in cases where ACI alone is not adequate (Fig. \ref{fig:Average_ACI_DL_Cn2_7e-15}(b1–c1)). This limitation arises because the receptive fields of conventional CNNs, even those with residual learning architectures such as DnCNN, are not well suited for capturing the large-scale geometric transformations or nonlinear phase distortions introduced by strong turbulence \cite{schuler2015learning}.

To overcome these limitations, we explored an alternative integration of ACI with the DL model. While ACI demonstrates potential in preserving geometric integrity and reducing deformation, its performance varies significantly with different atmospheric realizations, occasionally failing to restore the complete structure. This variability is illustrated in Fig. \ref{fig:Average_ACI_DL_Cn2_7e-15} (c1-c3). To address such deficiencies, a DL model specifically designed to handle geometric deformation could be beneficial. Additionally, since ACI-induced distortion varies across realizations, averaging multiple atmospheric realizations can further enhance geometric accuracy. In Fig. \ref{fig:Average_ACI_DL_Cn2_7e-15}, we present three target ground truths in column (a), their passive reconstructions in column (b), active reconstructions in column (c), averaged active reconstructions across ten realizations in column (d), and the DL applied to averaged reconstructions in column (e).

To thoroughly assess this methodology, we applied it across higher turbulence conditions. Fig. \ref{fig:Average_ACI_DL_2} shows ground truth targets in (a) and their reconstructions under three turbulence intensities ($C_n^2 = 0.7 \times 10^{-14}$, $1.7 \times 10^{-14}$, and $2.7 \times 10^{-14} \textrm{m}^{-2/3}$). Columns (b) to (e) illustrate passive reconstruction, passive with DL enhancement, averaged passive reconstructions, and averaged passive reconstructions with DL, respectively. Similarly, columns (f) to (i) show active reconstructions, active with DL, averaged active reconstructions, and averaged active reconstructions with DL enhancement. NCC values, denoted by $\xi$, quantify reconstruction fidelity relative to the ground truth.

\subsection{Alternative Deep Learning Models}

\begin{figure}[H]
\centering
\includegraphics[width=0.79\linewidth]{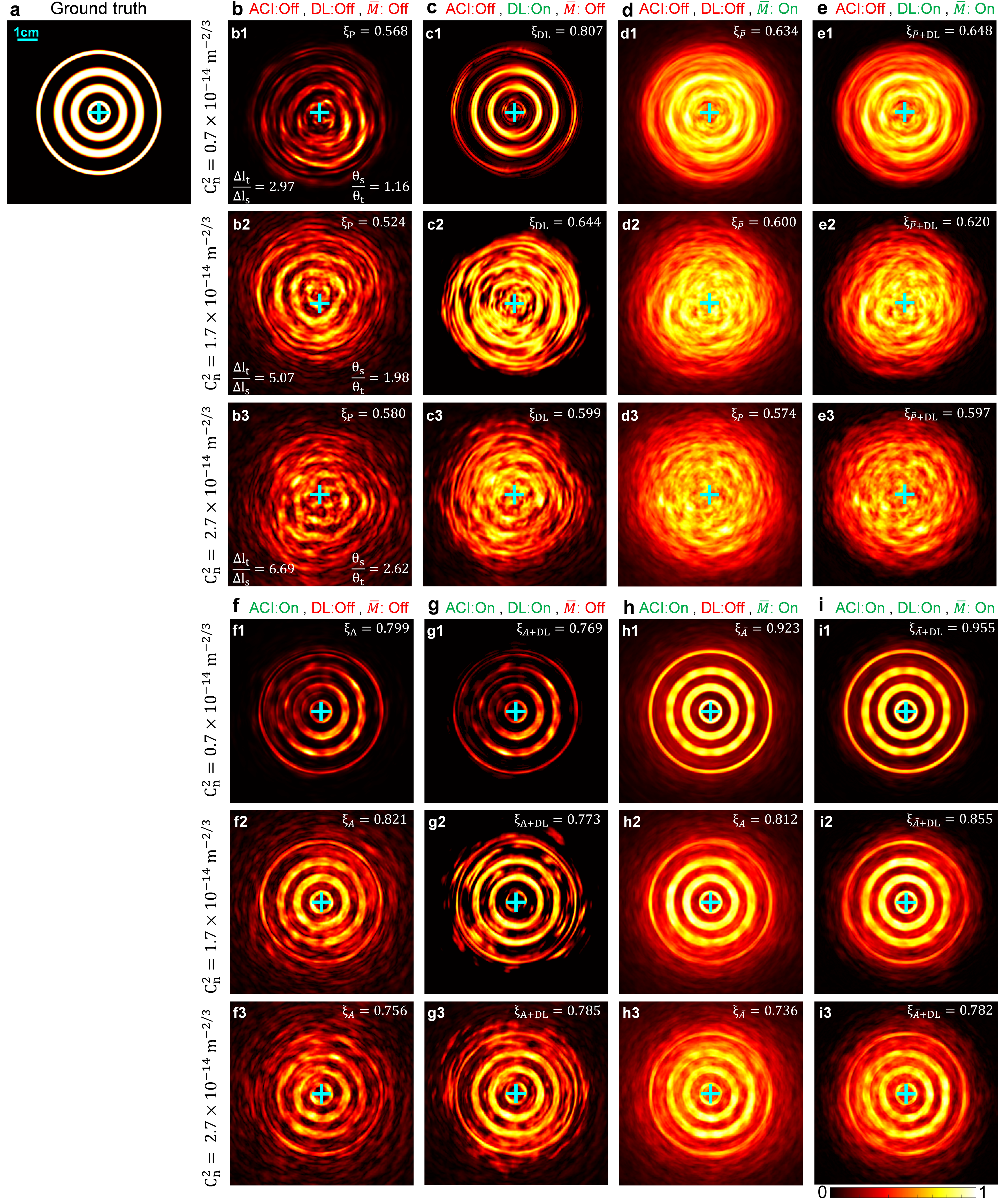}
\caption{A representative target is propagated through atmospheric turbulence and reconstructed under various scenarios across three turbulence levels ($C_n^2 = 0.7 \times 10^{-14}$, $1.7 \times 10^{-14}$, and $2.7 \times 10^{-14}\,\textrm{m}^{-2/3}$). (a) shows the ground truth target. Column (b) presents the target after propagation through turbulence and subsequent passive reconstruction. Column (c) shows the output of our DL model applied to the passive results in column (b). Column (d) presents passive reconstructions averaged over ten independent atmospheric realizations, and column (e) displays the DL-enhanced results applied to the averaged passive reconstructions. Columns (f) to (i) show reconstructions under active illumination: (f) active only, (g) active with DL, (h) averaged active reconstructions over ten realizations, and (i) averaged active reconstructions with DL enhancement. Turbulence distortion severity increases progressively from rows 1 to 3 and again from rows 4 to 6. The NCC between each reconstruction and the corresponding ground truth is denoted by $\xi$.}
\label{fig:Average_ACI_DL_2}
\end{figure}

Convolutional Neural Networks (CNNs) are foundational tools in image restoration, particularly effective in correcting low-order aberrations such as blur and scintillation. These networks excel at learning localized mappings between distorted and clean images \cite{dong2015image, mao2016image} and have been successfully applied to turbulence mitigation by modeling distortions as blur kernels \cite{cheng2023restoration}. However, their fixed kernel sizes and limited receptive fields constrain their ability to correct spatially varying geometric deformations characteristic of strong turbulence. The DL model employed in this study exhibits similar limitations. Since it lacks geometric awareness, arbitrary suppression of ripples and enhancement of contrast may lead to varying levels of reconstruction artifacts, as illustrated in Fig. \ref{fig:Average_ACI_DL_2}(g2–g3).
 
A DL model that adapts to turbulence severity, estimated via the refractive index structure parameter ($C_n^2$), may potentially perform better across conditions. Incorporating a confidence threshold or adaptive attention mechanism based on $C_n^2$ levels could improve model reliability under diverse conditions. Moreover, when integrated with ACI, a DL model should target residual distortions that ACI alone cannot fully suppress (Fig. \ref{fig:Average_ACI_DL_2}(f2–f3)). As optical modes traverse turbulent media, they encounter a combination of beam wandering, geometric deformation, contrast loss, and higher- and lower-order aberrations \cite{hill2025deep}. While ACI effectively mitigates beam wander and partially corrects geometric and contrast distortions, its performance degrades under severe turbulence, as evident in Fig. \ref{fig:Average_ACI_DL_2}(f1–f3). Therefore, the ideal complementary DL model should focus on residual geometric deformation and high-order aberrations, distortions that are spatially complex.

Standard CNNs often struggle to capture such complexities. To address this, advanced architectures such as deformable convolutional networks \cite{dai2017deformable} or spatial transformer networks \cite{jaderberg2015spatial} can be employed. These models explicitly learn spatial transformations and adapt to local geometric changes, offering enhanced resilience to severe distortions. This synergy has the potential to boost DL performance.

Note that while the present work explores the effectiveness of combining DnCNN with transfer learning for integration with ACI, it does not fully explore the breadth of possible enhancements achievable through this integration. Choices such as network architecture adjustments, alternative transfer strategies, different fine-tuning strategies, or dataset selection were beyond the scope of the present work and may reveal further improvements in enhancing ACI performance.

To determine viable alternatives to DnCNN for integration with ACI, a comparative analysis with state-of-the-art deep learning models developed for image restoration and enhancement is necessary and is left for future work. Generative Adversarial Networks (GANs) represent another promising direction, especially for restoring high-frequency details lost in turbulence. Through adversarial training, the generator learns to produce perceptually sharp images, while the discriminator distinguishes real from synthetic outputs \cite{ahmad2022new,yasarla2021learning}. However, this perceptual optimization comes with a critical drawback: the potential for hallucination—generating plausible yet inaccurate features. GANs, being trained for realism rather than fidelity, may fabricate or omit important information, particularly when exposed to out-of-distribution inputs \cite{bau2019seeing}.

This poses a serious risk in scientific and surveillance applications, where even subtle errors can lead to misinterpretation. For instance, GANs have been shown to hallucinate tumors in medical scans \cite{cohen2018distribution}. Although CycleGAN facilitates training with unpaired datasets \cite{zhu2017unpaired}, the interpretability and reliability of GAN-based outputs under real-world variability remain uncertain \cite{cohen2020limits}. A recent advancement, MS-TS-GAN \cite{lu2025research}, leverages multi-scale feature extraction and temporal information from sequential frames to improve restoration of turbulence-degraded images. By fusing spatial and temporal information, this model effectively addresses geometric distortions and high-frequency losses typical in atmospheric imaging scenarios. Their results highlight GANs’ potential for turbulence mitigation when temporal coherence is exploited.

U-Net architectures are highly effective for image-to-image translation tasks involving geometric distortions. Their encoder-decoder structure, enhanced by skip connections, captures both high-level semantic features and low-level spatial detail. U-Nets have shown success in learning dense flow fields to "unwarp" distorted images and have been applied in video stabilization and object tracking \cite{vint2020analysis,ronneberger2015u,zhao2023fast}. Their ability to handle spatially varying deformations makes them strong candidates for turbulence correction.

To address dynamic, temporally evolving turbulence, models that capture temporal dependencies—such as Recurrent Neural Networks (RNNs), Long Short-Term Memory (LSTM) networks, and Vision Transformers (ViTs)—are particularly well-suited. These models process sequences of distorted frames to correct temporal distortions like beam wander and tip/tilt jitter. Temporal filtering architectures such as ConvLSTMs have demonstrated effectiveness in learning distortion dynamics over time, enabling smoother and more consistent reconstructions \cite{gao2019atmospheric, peng2018red}. Other architectures such as SwinIR \cite{liang2021swinir}, EDSR \cite{lim2017enhanced}, and Restormer \cite{zamir2022restormer} are also strong candidates for a comparative study. While DnCNN provides a well-understood and computationally efficient baseline, these newer models offer increased representational capacity and improved restoration fidelity, particularly under complex distortion regimes.

Additionally, DL model performance strongly depends on the dataset. Although ACI tolerates variability in object shapes \cite{ghoshroy2024enhancing, moazzam2025remote, moazzam2023active}, DL models generally require datasets that are comprehensive and representative. In the context of atmospheric imaging, realism in data generation is equally crucial. Models trained solely on synthetic data may fail to generalize if the turbulence statistics or target complexity deviate from real-world conditions. Therefore, domain adaptation techniques or physics-informed neural networks could bridge the domain gap between simulated and real turbulent data \cite{karniadakis2021physics, zhu2017unpaired}. Additionally, incorporating priors such as turbulence power spectra or Kolmogorov models during training could further ground the learning process in physically meaningful constraints \cite{safonov2025physically}.

Expanding the size and diversity of training datasets can also significantly improve DL model performance and strengthen its synergy with ACI. This is in line with recent findings in deep learning generalization theory, which emphasize that model capacity must be appropriately matched with both task complexity and data diversity to achieve effective transfer learning and scalable performance \cite{kaplan2020scaling, zhai2022scaling}.

\subsection{Deep Learning for CIPTs}

\begin{figure*}[th]
\centering
\includegraphics[width=0.75\linewidth]{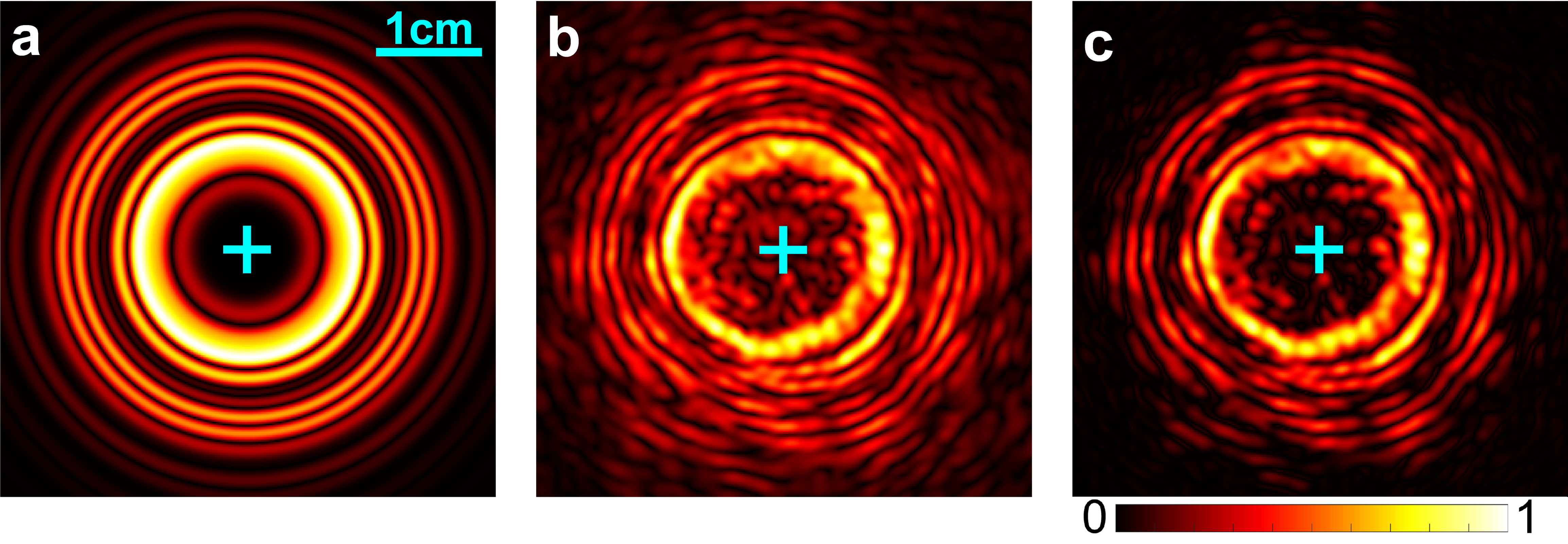}
\caption{Illustration of DL implementation with ACI partial targets. (a) An example partial target after propagation through a non-turbulent medium. (b) Reconstruction of the
same partial target obtained after correlation injection and subsequent propagation through
turbulence ($C_n^2 = 1.7 \times 10^{-14}\,\textrm{m}^{-2/3}$). (c) When applied to (b), the DL model and transfer learning strategy perform suboptimally. Although a degree of contrast sharpening is observed, geometric enhancement is largely absent.}
\label{fig:CIPT_1}
\end{figure*}

We previously described two DL implementations: one applied directly to passive and active reconstructed images, and another applied to averaged targets, which is viable only in active scenarios. The passive averaged reconstructions remain significantly distorted due to persistent beam wandering across realizations, rendering averaging ineffective for improvement. In the implementation of ACI, correlation-injected partial targets (CIPTs) are generated, giving a superposition of the target with an auxiliary source, each CIPT carrying partial yet high-fidelity target information through turbulent media \cite{ghoshroy2024enhancing}. Fig. \ref{fig:CIPT_1}(a) shows a partial target after propagation through a non-turbulent medium. Fig. \ref{fig:CIPT_1}(b) illustrates the reconstruction of the same partial target obtained after correlation injection (i.e., CIPT) and subsequent propagation through turbulence ($C_n^2 = 1.7 \times 10^{-14}\,\textrm{m}^{-2/3}$). Fig. \ref{fig:CIPT_1}(c) presents the effect of applying our DL model to Fig. \ref{fig:CIPT_1}(b). The underlying training dataset was the same as in Fig. \ref{fig:Average_ACI_DL_2}(g2) (i.e., no partial targets were used). A comparison between Figs. \ref{fig:CIPT_1}(a) and (b) indicates that there is considerable room for enhancement in the reconstructed partial target. However, the selected DL model and transfer learning strategy underperform as seen in Fig. \ref{fig:CIPT_1}(c). In principle, the performance seen in Fig. \ref{fig:CIPT_1}(c) can be improved if the transfer learning fine-tuning strategy is optimized, for example, by carefully selecting the partial target dataset with appropriate diversity or learning rate adjustment. Therefore, DL implementation with ACI partial targets presents a unique opportunity to improve image reconstruction beyond what is traditionally possible.       

CIPTs experience a smaller diffraction cone compared to the full target, resulting in reduced distortion and making them more amenable to enhancement by DL models. Improving the quality of CIPTs can, in turn, facilitate more accurate reconstruction of the full target by reinforcing high-fidelity structural components. When combined with the spatially and temporally adaptive architectures discussed above, this framework offers strong potential for achieving robust and precise reconstructions under a wide range of turbulent conditions.

Despite the advantages offered by ACI, CIPTs still exhibit geometric deformation and residual higher- and lower-order aberrations. The current DL model, lacking sensitivity to geometric structures, is not well-equipped to correct these complex distortions, limiting its effectiveness when applied to CIPT inputs. This highlights the need for alternative architectures specifically tailored to the unique characteristics of CIPTs.

An effective DL model integrated with ACI should target the residual distortion components that ACI alone cannot fully mitigate. As optical modes propagate through turbulent media, they encounter a combination of beam wander, geometric distortion, contrast degradation, and both low-order and high-order aberrations \cite{hill2025deep}. While ACI effectively reduces beam wander and partially restores geometric and contrast fidelity, its performance declines under severe turbulence. We believe that a complementary DL model should focus on addressing the remaining geometric distortions and nonlinear, spatially varying aberrations, features that standard CNNs often fail to capture due to their limited receptive fields.

Emerging architectures such as deformable convolutional networks \cite{dai2017deformable} and spatial transformer networks \cite{jaderberg2015spatial} present promising alternatives. These models explicitly learn spatial transformations and dynamically adapt to local geometric variability, making them well-suited for correcting residual distortions in CIPTs. Given that CIPTs experience only partial atmospheric distortion due to their narrower diffraction cone, incorporating deformation-aware mechanisms can enhance their structural fidelity and ultimately improve the reconstruction quality of the full target when used in conjunction with ACI.

\section{Conclusion}

The integration of ACI with DL models presents a compelling path forward for robust optical beam propagation through turbulence. ACI introduces physics-informed distortion compensation by injecting structural correlations directly into the illumination field, offering propagation stability in challenging propagation regimes. However, this stability can be limited by sensitivity to initial characterization errors and a lack of real-time adaptability under dynamic or anisoplanatic conditions. DL, by contrast, provides flexible, data-driven tools that excel at modeling nonlinear distortions and extracting latent structure from degraded observations \cite{mccann2017convolutional}. This complementarity positions ACI and DL as natural partners: ACI conditions the signal for robustness, while DL offers adaptive post-processing or correction strategies that respond to evolving or deteriorating turbulence conditions.

ACI alone substantially improves structural integrity by mitigating beam wander and geometric deformation, but its effectiveness diminishes under severe turbulence \cite{ghoshroy2024enhancing, moazzam2025remote}. Our DnCNN-based transfer-learning model reduces residual distortions, such as ripple artifacts and contrast loss, yet is less effective at correcting strong geometric deformations and higher-order aberrations. Together, these results demonstrate the feasibility of this hybrid approach while underscoring the need for architectures with stronger geometric awareness for this application.

This reveals that several avenues for integration are feasible. Advanced DL models such as U-Net \cite{ronneberger2015u}, SwinIR \cite{liang2021swinir}, EDSR \cite{lim2017enhanced}, and Restormer \cite{zamir2022restormer} may be used to enhance ACI-transmitted fields in a post-detection configuration, recovering spatial fidelity lost during propagation. Alternatively, DL architectures can be incorporated into end-to-end learning frameworks, where the ACI parameters and the network weights are jointly optimized to minimize reconstruction artifacts
. Transfer learning further enables adaptation of pretrained networks to ACI-specific distortions, reducing data requirements while preserving generalization \cite{jebur2024comprehensive}. Metrics such as PSNR and SSIM provide quantitative benchmarks for reconstruction fidelity, while no-reference quality measures like  NIQE \cite{mittal2012making}, and BRISQUE \cite{mittal2012no} may allow for robust evaluation in realistic, feedback-limited conditions. Importantly, architectural analysis of newer models can help identify design principles, such as multi-scale aggregation, residual feedback, or transformer-based attention, that may be well suited to the structured priors introduced by ACI.

We additionally evaluated DL performance on averages across multiple atmospheric realizations. This strategy improved consistency and robustness by smoothing stochastic distortions and further enhancing reconstruction fidelity. Averaging was particularly beneficial in active scenarios, where reduced beam wander across realizations improved alignment, but it offered little value in passive setups with persistent structural misregistration.

CIPTs propagate through a narrower diffraction cone and typically incur less distortion. Therefore, their direct enhancement with DL may be potentially more effective on these targets. However, optimal integration will likely require DL architectures explicitly designed to handle residual geometric and high-order aberrations that ACI does not fully correct; standard CNNs, including our current model, lack sufficient geometric priors and can introduce artifacts or fail to recover structure accurately.

Looking ahead, this hybrid approach opens up exciting opportunities for the development of next-generation optical systems. For instance, differentiable simulation models \cite{newbury2024review} for the ACI framework could be embedded into neural network training loops, enabling fully physics-informed learning pipelines \cite{chakrabarti2016neural,andrychowicz2016learning}. Reinforcement learning or Bayesian optimization may be used to actively tune ACI parameters in real time based on image quality feedback \cite{parvizi2023reinforcement,nousiainen2021adaptive}. Closed-loop systems combining fast DL inference with adaptive illumination could enable robust imaging, tracking, or communication over long distances and in dynamically changing environments. Overall, the convergence of physics-driven correction frameworks like ACI and intelligent algorithmic correction represents a powerful avenue for overcoming the limitations of purely passive or reactive strategies in turbulent propagation regimes.

\section*{Disclosures}
The authors declare no financial interests, commercial affiliations, or other potential conflicts of interest that could have influenced the objectivity of this research or the preparation of this manuscript.

\section*{Data Availability}
All data and code used in this study are available from the corresponding author upon request.

\section*{Acknowledgement}
This work was supported by the National Science Foundation (EAR-2221730). 

\bibliographystyle{unsrt} 
\bibliography{references}

@article{bertolotti2022imaging,
  title={Imaging in complex media},
  author={Bertolotti, Jacopo and Katz, Ori},
  journal={Nature Physics},
  volume={18},
  number={9},
  pages={1008--1017},
  year={2022},
  publisher={Nature Publishing Group UK London}
}

@inproceedings{harmeling2009online,
  title={Online blind deconvolution for astronomical imaging},
  author={Harmeling, Stefan and Hirsch, Michael and Sra, Suvrit and Sch{\"o}lkopf, Bernhard},
  booktitle={Proceedings of the IEEE International Conference on Computational Photography},
  pages={1--7},
  year={2009},
  organization={IEEE}
}

@book{campisi2017blind,
  title={Blind image deconvolution: theory and applications},
  author={Campisi, Patrizio and Egiazarian, Karen},
  year={2017},
  publisher={CRC press}
}

@article{hampson2021adaptive,
  title={Adaptive optics for high-resolution imaging},
  author={Hampson, Karen M and Turcotte, Rapha{\"e}l and Miller, Donald T and Kurokawa, Kazuhiro and Males, Jared R and Ji, Na and Booth, Martin J},
  journal={Nature Reviews Methods Primers},
  volume={1},
  number={1},
  pages={68},
  year={2021},
  publisher={Nature Publishing Group UK London}
}

@inproceedings{schuler2013machine,
  title={A machine learning approach for non-blind image deconvolution},
  author={Schuler, Christian J and Christopher Burger, Harold and Harmeling, Stefan and Scholkopf, Bernhard},
  booktitle={Proceedings of the IEEE conference on computer vision and pattern recognition},
  pages={1067--1074},
  year={2013}
}

@article{levin2011understanding,
  title={Understanding blind deconvolution algorithms},
  author={Levin, Anat and Weiss, Yair and Durand, Fredo and Freeman, William T},
  journal={IEEE transactions on pattern analysis and machine intelligence},
  volume={33},
  number={12},
  pages={2354--2367},
  year={2011},
  publisher={IEEE}
}

@article{lohani2018turbulence,
  title={Turbulence correction with artificial neural networks},
  author={Lohani, Sanjaya and Glasser, Ryan T},
  journal={Optics letters},
  volume={43},
  number={11},
  pages={2611--2614},
  year={2018},
  publisher={Optica Publishing Group}
}

@article{feng2022turbugan,
  title={Turbugan: An adversarial learning approach to spatially-varying multiframe blind deconvolution with applications to imaging through turbulence},
  author={Feng, Brandon Y and Xie, Mingyang and Metzler, Christopher A},
  journal={IEEE Journal on Selected Areas in Information Theory},
  volume={3},
  number={3},
  pages={543--556},
  year={2022},
  publisher={IEEE}
}

@article{ghoshroy2024enhancing,
  title={Enhancing Complex Light Beam Propagation in Turbulent Atmosphere with Active Convolved Illumination},
  author={Ghoshroy, Anindya and Davis, James and Moazzam, Adrian A and Askari, Roohollah and G{\"u}ney, Durdu {\"O}},
  journal={ACS Photonics},
  volume={11},
  pages={4541-4558},
  year={2024},
  publisher={ACS Publications}
}

@inproceedings{miller2021machine,
  title={A machine learning approach to improving quality of atmospheric turbulence simulation},
  author={Miller, Kevin J and Du Bosq, Todd},
  booktitle={Infrared Imaging Systems: Design, Analysis, Modeling, and Testing XXXII},
  volume={11740},
  pages={126--138},
  year={2021},
  organization={SPIE}
}

@article{zhang2024imaging,
  title={Imaging through the atmosphere using turbulence mitigation transformer},
  author={Zhang, Xingguang and Mao, Zhiyuan and Chimitt, Nicholas and Chan, Stanley H},
  journal={IEEE Transactions on Computational Imaging},
  year={2024},
  publisher={IEEE}
}

@article{gao2019atmospheric,
  title={Atmospheric turbulence removal using convolutional neural network},
  author={Gao, Jing and Anantrasirichai, Nantheera and Bull, David},
  journal={arXiv preprint arXiv:1912.11350},
  year={2019}
}

@article{schmidt2010numerical,
  title={Numerical simulation of optical wave propagation with examples in MATLAB},
  author={Schmidt, Jason D},
  journal={(No Title)},
  year={2010},
  publisher={Spie}
}

@book{ke2023generation,
  title={Generation, transmission, detection, and application of vortex beams},
  author={Ke, Xizheng and Wang, J},
  year={2023},
  publisher={Springer}
}

@article{liu2019deep,
  title={Deep learning based atmospheric turbulence compensation for orbital angular momentum beam distortion and communication},
  author={Liu, Junmin and Wang, Peipei and Zhang, Xiaoke and He, Yanliang and Zhou, Xinxing and Ye, Huapeng and Li, Ying and Xu, Shixiang and Chen, Shuqing and Fan, Dianyuan},
  journal={Optics express},
  volume={27},
  number={12},
  pages={16671--16688},
  year={2019},
  publisher={Optica Publishing Group}
}

@article{zhai2020turbulence,
  title={Turbulence aberration correction for vector vortex beams using deep neural networks on experimental data},
  author={Zhai, Yanwang and Fu, Shiyao and Zhang, Jianqiang and Liu, Xueting and Zhou, Heng and Gao, Chunqing},
  journal={Optics Express},
  volume={28},
  number={5},
  pages={7515--7527},
  year={2020},
  publisher={Optica Publishing Group}
}

@inproceedings{yasarla2021learning,
  title={Learning to restore images degraded by atmospheric turbulence using uncertainty},
  author={Yasarla, Rajeev and Patel, Vishal M},
  booktitle={2021 IEEE International Conference on Image Processing (ICIP)},
  pages={1694--1698},
  year={2021},
  organization={IEEE}
}

@article{rai2022removing,
  title={Removing atmospheric turbulence via deep adversarial learning},
  author={Rai, Shyam Nandan and Jawahar, CV},
  journal={IEEE Transactions on Image Processing},
  volume={31},
  pages={2633--2646},
  year={2022},
  publisher={IEEE}
}

@inproceedings{jaiswal2023physics,
  title={Physics-driven turbulence image restoration with stochastic refinement},
  author={Jaiswal, Ajay and Zhang, Xingguang and Chan, Stanley H and Wang, Zhangyang},
  booktitle={Proceedings of the IEEE/CVF International Conference on Computer Vision},
  pages={12170--12181},
  year={2023}
}

@article{chen2019u,
  title={U-net like deep autoencoders for deblurring atmospheric turbulence},
  author={Chen, Gongping and Gao, Zhisheng and Wang, Qiaolu and Luo, Qingqing},
  journal={Journal of Electronic Imaging},
  volume={28},
  number={5},
  pages={053024--053024},
  year={2019},
  publisher={Society of Photo-Optical Instrumentation Engineers}
}

@inproceedings{repasi2011computer,
  title={Computer simulation of image degradations by atmospheric turbulence for horizontal views},
  author={Repasi, Endre and Weiss, Robert},
  booktitle={Infrared Imaging Systems: Design, Analysis, Modeling, and Testing XXII},
  volume={8014},
  pages={279--287},
  year={2011},
  organization={SPIE}
}

@inproceedings{schwartzman2017turbulence,
  title={Turbulence-induced 2d correlated image distortion},
  author={Schwartzman, Armin and Alterman, Marina and Zamir, Rotem and Schechner, Yoav Y},
  booktitle={Proceedings of the IEEE International Conference on Computational Photography},
  pages={1--13},
  year={2017},
  organization={IEEE}
}

@inproceedings{miller2019data,
  title={A data-constrained algorithm for the emulation of long-range turbulence-degraded video},
  author={Miller, Kevin J and Preece, Bradley and Du Bosq, Todd W and Leonard, Kevin R},
  booktitle={Infrared Imaging Systems: Design, Analysis, Modeling, and Testing XXX},
  volume={11001},
  pages={204--214},
  year={2019},
  organization={SPIE}
}

@article{zhang2017beyond,
  title={Beyond a gaussian denoiser: Residual learning of deep cnn for image denoising},
  author={Zhang, Kai and Zuo, Wangmeng and Chen, Yunjin and Meng, Deyu and Zhang, Lei},
  journal={IEEE transactions on image processing},
  volume={26},
  number={7},
  pages={3142--3155},
  year={2017},
  publisher={IEEE}
}

@inproceedings{zamir2022restormer,
  title={Restormer: Efficient transformer for high-resolution image restoration},
  author={Zamir, Syed Waqas and Arora, Aditya and Khan, Salman and Hayat, Munawar and Khan, Fahad Shahbaz and Yang, Ming-Hsuan},
  booktitle={Proceedings of the IEEE/CVF conference on computer vision and pattern recognition},
  pages={5728--5739},
  year={2022}
}

@article{guo2022blind,
  title={Blind restoration of images distorted by atmospheric turbulence based on deep transfer learning},
  author={Guo, Yiming and Wu, Xiaoqing and Qing, Chun and Su, Changdong and Yang, Qike and Wang, Zhiyuan},
  journal={Photonics},
  volume={9},
  number={8},
  pages={582},
  year={2022},
  publisher={MDPI}
}

@inproceedings{torralba2011unbiased,
  title={Unbiased look at dataset bias},
  author={Torralba, Antonio and Efros, Alexei A},
  booktitle={Proceedings of the IEEE Conference on Computer Vision and Pattern Recognition},
pages={1521--1528},
  year={2011},
  organization={IEEE}
}

@inproceedings{sun2017revisiting,
  title={Revisiting unreasonable effectiveness of data in deep learning era},
  author={Sun, Chen and Shrivastava, Abhinav and Singh, Saurabh and Gupta, Abhinav},
  booktitle={Proceedings of the IEEE International Conference on Computer Vision},
  pages={843--852},
  year={2017}
}

@article{he2009learning,
  title={Learning from imbalanced data},
  author={He, Haibo and Garcia, Edwardo A},
  journal={IEEE Transactions on knowledge and data engineering},
  volume={21},
  number={9},
  pages={1263--1284},
  year={2009},
  publisher={Ieee}
}

@article{buda2018systematic,
  title={A systematic study of the class imbalance problem in convolutional neural networks},
  author={Buda, Mateusz and Maki, Atsuto and Mazurowski, Maciej A},
  journal={Neural networks},
  volume={106},
  pages={249--259},
  year={2018},
  publisher={Elsevier}
}

@article{yosinski2014transferable,
  title={How transferable are features in deep neural networks?},
  author={Yosinski, Jason and Clune, Jeff and Bengio, Yoshua and Lipson, Hod},
  journal={Advances in neural information processing systems},
  volume={27},
  year={2014}
}

@article{sadatgol2015plasmon,
  title={Plasmon injection to compensate and control losses in negative index metamaterials},
  author={Sadatgol, Mehdi and {\"O}zdemir, {\c{S}}ahin K and Yang, Lan and G{\"u}ney, Durdu {\"O}},
  journal={Physical review letters},
  volume={115},
  number={3},
  pages={035502},
  year={2015},
  publisher={APS}
}

@article{ghoshroy2020loss,
  title={Loss compensation in metamaterials and plasmonics with virtual gain},
  author={Ghoshroy, Anindya and {\"O}zdemir, {\c{S}}ahin K and G{\"u}ney, Durdu {\"O}},
  journal={Optical Materials Express},
  volume={10},
  number={8},
  pages={1862--1880},
  year={2020},
  publisher={Optical Society of America}
}

@article{adams2016review,
  title={Review of near-field optics and superlenses for sub-diffraction-limited nano-imaging},
  author={Adams, Wyatt and Sadatgol, Mehdi and G{\"u}ney, Durdu {\"O}},
  journal={AIP Advances},
  volume={6},
  number={10},
  year={2016},
  publisher={AIP Publishing}
}

@article{adams2016bringing,
  title={Bringing the ‘perfect lens’ into focus by near-perfect compensation of losses without gain media},
  author={Adams, Wyatt and Sadatgol, Mehdi and Zhang, Xu and G{\"u}ney, Durdu {\"O}},
  journal={New Journal of Physics},
  volume={18},
  number={12},
  pages={125004},
  year={2016},
  publisher={IOP Publishing}
}

@article{zhang2016enhancing,
  title={Enhancing the resolution of hyperlens by the compensation of losses without gain media},
  author={Zhang, Xu and Adams, Wyatt and Sadatgol, Mehdi and G{\"u}ney, Durdu {\"O}},
  journal={arXiv preprint arXiv:1607.07466},
  year={2016}
}

@article{zhang2017analytical,
  title={Analytical description of inverse filter emulating the plasmon injection loss compensation scheme and implementation for ultrahigh-resolution hyperlens},
  author={Zhang, Xu and Adams, Wyatt and G{\"u}ney, Durdu {\"O}},
  journal={JOSA B},
  volume={34},
  number={6},
  pages={1310--1318},
  year={2017},
  publisher={Optica Publishing Group}
}

@article{adams2017plasmonic,
  title={Plasmonic superlens image reconstruction using intensity data and equivalence to structured light illumination for compensation of losses},
  author={Adams, Wyatt and Ghoshroy, Anindya and G{\"u}ney, Durdu {\"O}},
  journal={JOSA B},
  volume={34},
  number={10},
  pages={2161--2168},
  year={2017},
  publisher={Optica Publishing Group}
}

@article{adams2018plasmonic,
  title={Plasmonic superlens imaging enhanced by incoherent active convolved illumination},
  author={Adams, Wyatt and Ghoshroy, Anindya and G{\"u}ney, Durdu {\"O}},
  journal={ACS Photonics},
  volume={5},
  number={4},
  pages={1294--1302},
  year={2018},
  publisher={ACS Publications}
}

@article{adams2022incoherent,
  title={Incoherent Active Convolved Illumination Enhances the Signal-to-Noise Ratio for Shot Noise: Experimental Evidence},
  author={Adams, Wyatt and Ghoshroy, Anindya and G{\"u}ney, Durdu {\"O}},
  journal={Physical Review Applied},
  volume={18},
  number={6},
  pages={064080},
  year={2022},
  publisher={APS}
}

@article{ghoshroy2020theory,
  title={Theory of coherent active convolved illumination for superresolution enhancement},
  author={Ghoshroy, Anindya and Adams, Wyatt and G{\"u}ney, Durdu {\"O}},
  journal={JOSA B},
  volume={37},
  number={8},
  pages={2452--2463},
  year={2020},
  publisher={Optica Publishing Group}
}

@inproceedings{moazzam2023active,
  title={Active convolved illumination towards enhanced remote sensing through a turbulent atmosphere},
  author={Moazzam, A. A. and Ghoshroy, A. and Askari, R. and G{\"u}ney, Durdu {\"O}},
  booktitle={AGU 2023, 11-15 December 2023, San Francisco, CA, USA.},
  pages={S11F-0322},
  year={2023},
}

@article{wang2004image,
  title={Image quality assessment: from error visibility to structural similarity},
  author={Wang, Zhou and Bovik, Alan C and Sheikh, Hamid R and Simoncelli, Eero P},
  journal={IEEE transactions on image processing},
  volume={13},
  number={4},
  pages={600--612},
  year={2004},
  publisher={IEEE}
}

@article{zhao2016loss,
  title={Loss functions for image restoration with neural networks},
  author={Zhao, Hang and Gallo, Orazio and Frosio, Iuri and Kautz, Jan},
  journal={IEEE Transactions on computational imaging},
  volume={3},
  number={1},
  pages={47--57},
  year={2016},
  publisher={IEEE}
}

@article{kaplan2020scaling,
  title={Scaling laws for neural language models},
  author={Kaplan, Jared and McCandlish, Sam and Henighan, Tom and Brown, Tom B and Chess, Benjamin and Child, Rewon and Gray, Scott and Radford, Alec and Wu, Jeffrey and Amodei, Dario},
  journal={arXiv preprint arXiv:2001.08361},
  year={2020}
}

@article{warhaft2002turbulence,
  title={Turbulence in nature and in the laboratory},
  author={Warhaft, Z},
  journal={Proceedings of the National Academy of Sciences},
  volume={99},
  number={suppl\_1},
  pages={2481--2486},
  year={2002},
  publisher={National Academy of Sciences}
}

@book{roggemann2018imaging,
  title={Imaging through turbulence},
  author={Roggemann, Michael C and Welsh, Byron M},
  year={2018},
  publisher={CRC press}
}

@article{andrews2005laser,
  title={Laser beam propagation through random media},
  author={Andrews, Larry C and Phillips, Ronald L},
  journal={Laser Beam Propagation Through Random Media: Second Edition},
  year={2005}
}

@book{hardy1998adaptive,
  title={Adaptive optics for astronomical telescopes},
  author={Hardy, John W},
  volume={16},
  year={1998},
  publisher={Oxford university press}
}

@article{fried1966optical,
  title={Optical resolution through a randomly inhomogeneous medium for very long and very short exposures},
  author={Fried, David L},
  journal={Journal of the Optical Society of America},
  volume={56},
  number={10},
  pages={1372--1379},
  year={1966},
  publisher={Optical Society of America}
}

@article{tatarskii1971effects,
  title={The effects of the turbulent atmosphere on wave propagation},
  author={Tatarskii, Valerian Ilitch},
  journal={Jerusalem: Israel Program for Scientific Translations},
  pages={472},
  year={1971}
}

@incollection{roddier1981v,
  title={The effects of atmospheric turbulence in optical astronomy},
  author={Roddier, Fran{\c{c}}ois},
  booktitle={Progress in optics},
  chapter={V},
  volume={19},
  pages={281--376},
  year={1981},
  publisher={Elsevier}
}

@article{cox2020structured,
  title={Structured light in turbulence},
  author={Cox, Mitchell A and Mphuthi, Nokwazi and Nape, Isaac and Mashaba, Nikiwe and Cheng, Ling and Forbes, Andrew},
  journal={IEEE Journal of Selected Topics in Quantum Electronics},
  volume={27},
  number={2},
  pages={1--21},
  year={2020},
  publisher={IEEE}
}

@article{wang2018performance,
  title={Performance analysis of an adaptive optics system for free-space optics communication through atmospheric turbulence},
  author={Wang, Yukun and Xu, Huanyu and Li, Dayu and Wang, Rui and Jin, Chengbin and Yin, Xianghui and Gao, Shijie and Mu, Quanquan and Xuan, Li and Cao, Zhaoliang},
  journal={Scientific reports},
  volume={8},
  number={1},
  pages={1124},
  year={2018},
  publisher={Nature Publishing Group UK London}
}

@article{lv2025adaptive,
  title={An adaptive lucky imaging method for turbulence-degraded image restoration},
  author={Lv, Pin and Shi, Tiezhu and Den, Dongping and Wang, Mengdi and Liu, Qian and Wu, Guofeng},
  journal={IET Image Processing},
  volume={19},
  number={1},
  pages={e13312},
  year={2025},
  publisher={Wiley Online Library}
}

@article{fried1978probability,
  title={Probability of getting a lucky short-exposure image through turbulence},
  author={Fried, David L},
  journal={Journal of the Optical Society of America},
  volume={68},
  number={12},
  pages={1651--1658},
  year={1978},
  publisher={Optical Society of America}
}

@inproceedings{zhang2011astronomical,
  title={Astronomical image restoration through atmosphere turbulence by lucky imaging},
  author={Zhang, Shixue and Wu, Yuanhao and Zhao, Jinyu and Wang, Jianli},
  booktitle={Third International Conference on Digital Image Processing},
  volume={8009},
  pages={70--74},
  year={2011},
  organization={SPIE}
}

@article{furhad2016restoring,
  title={Restoring atmospheric-turbulence-degraded images},
  author={Furhad, Md Hasan and Tahtali, Murat and Lambert, Andrew},
  journal={Applied optics},
  volume={55},
  number={19},
  pages={5082--5090},
  year={2016},
  publisher={Optical Society of America}
}

@inproceedings{zhu2010image,
  title={Image reconstruction from videos distorted by atmospheric turbulence},
  author={Zhu, Xiang and Milanfar, Peyman},
  booktitle={Visual Information Processing and Communication},
  volume={7543},
  pages={228--235},
  year={2010},
  organization={SPIE}
}

@inproceedings{snell2017learning,
  title={Learning to generate images with perceptual similarity metrics},
  author={Snell, Jake and Ridgeway, Karl and Liao, Renjie and Roads, Brett D and Mozer, Michael C and Zemel, Richard S},
  booktitle={Proceedings of the IEEE international conference on image processing},
  pages={4277--4281},
  year={2017},
  organization={IEEE}
}

@article{hill2025deep,
  title={Deep learning techniques for atmospheric turbulence removal: a review},
  author={Hill, Paul and Anantrasirichai, Nantheera and Achim, Alin and Bull, David},
  journal={Artificial Intelligence Review},
  volume={58},
  number={4},
  pages={101},
  year={2025},
  publisher={Springer}
}

@article{moazzam2025remote,
  title={Remote Sensing of Seismic Signals via Enhanced Moir{\'e}-Based Apparatus Integrated with Active Convolved Illumination},
  author={Moazzam, Adrian A and Ghoshroy, Anindya and G{\"u}ney, Durdu {\"O} and Askari, Roohollah},
  journal={Remote Sensing},
  volume={17},
  number={12},
  pages={2032},
  year={2025},
  publisher={MDPI}
}

@article{schuler2015learning,
  title={Learning to deblur},
  author={Schuler, Christian J and Hirsch, Michael and Harmeling, Stefan and Sch{\"o}lkopf, Bernhard},
  journal={IEEE transactions on pattern analysis and machine intelligence},
  volume={38},
  number={7},
  pages={1439--1451},
  year={2015},
  publisher={IEEE}
}

@inproceedings{jaderberg2015spatial,
  author={Jaderberg, Max and Simonyan, Karen and Zisserman, Andrew and Kavukcuoglu, Koray},
  title={Spatial Transformer Networks},
  booktitle={Advances in Neural Information Processing Systems (NeurIPS)},
  volume={28},
  year={2015}
}

@inproceedings{dai2017deformable,
  title={Deformable convolutional networks},
  author={Dai, Jifeng and Qi, Haozhi and Xiong, Yuwen and Li, Yi and Zhang, Guodong and Hu, Han and Wei, Yichen},
  booktitle={Proceedings of the IEEE International Conference on Computer Vision},
  pages={764--773},
  year={2017}
}

@inproceedings{zhai2022scaling,
  title={Scaling Vision Transformers},
  author={Zhai, Xiaohua and Kolesnikov, Alexander and Houlsby, Neil and Beyer, Lucas},
  booktitle={Proceedings of the IEEE/CVF Conference on Computer Vision and Pattern Recognition},
  pages={12104--12113},
  year={2022},
  publisher={IEEE}
}

@article{karniadakis2021physics,
  title={Physics-informed machine learning},
  author={Karniadakis, George Em and Kevrekidis, Ioannis G and Lu, Lu and Perdikaris, Paris and Wang, Sifan and Yang, Liu},
  journal={Nature Reviews Physics},
  volume={3},
  number={6},
  pages={422--440},
  year={2021},
  publisher={Springer Nature},
}

@inproceedings{zhu2017unpaired,
  title={Unpaired Image-to-Image Translation Using Cycle-Consistent Adversarial Networks},
  author={Zhu, Jun-Yan and Park, Taesung and Isola, Phillip and Efros, Alexei A},
  booktitle={Proceedings of the IEEE International Conference on Computer Vision},
  pages={2223--2232},
  year={2017}
}

@article{safonov2025physically,
  title={Physically-informed neural networks for estimation of atmospheric optical turbulence intensity},
  author={Safonov, D A and Sinyutin, Y V and Shcheglov, D A},
  journal={Izvestiya Vysshikh Uchebnykh Zavedenii Fizika},
  volume={68},
  number={5},
  pages={66--73},
  year={2025},
}

@article{dong2015image,
  author={Dong, Chao and Loy, Chen Change and He, Kaiming and Tang, Xiaoou},
  title={Image Super-Resolution Using Deep Convolutional Networks},
  journal={IEEE Transactions on Pattern Analysis and Machine Intelligence},
  volume={38},
  number={2},
  pages={295--307},
  year={2015},
  publisher={IEEE}
}

@inproceedings{mao2016image,
  author={Mao, Xiaojiao and Shen, Chunhua and Yang, Yu-Bin},
  title={Image Restoration Using Very Deep Convolutional Encoder-Decoder Networks with Symmetric Skip Connections},
  booktitle={Advances in Neural Information Processing Systems},
  volume={29},
  year={2016}
}

@inproceedings{bau2019seeing,
  title={Seeing What a GAN Cannot Generate},
  author={Bau, David and Zhu, Jun-Yan and Wulff, Jonas and Peebles, William and Strobelt, Hendrik and Zhou, Bolei and Torralba, Antonio},
  booktitle={Proceedings of the IEEE/CVF International Conference on Computer Vision},
  pages={4502--4511},
  year={2019}
}

@inproceedings{cohen2020limits,
  title={On the limits of cross-domain generalization in automated X-ray prediction},
  author={Cohen, Joseph Paul and Hashir, Mohammad and Brooks, Rupert and Bertrand, Hadrien},
  booktitle={Medical Imaging with Deep Learning},
  pages={136--155},
  year={2020},
  organization={PMLR}
}

@article{peng2018red,
  title={Red-net: A recurrent encoder--decoder network for video-based face alignment},
  author={Peng, Xi and Feris, Rogerio S and Wang, Xiaoyu and Metaxas, Dimitris N},
  journal={International Journal of Computer Vision},
  volume={126},
  number={10},
  pages={1103--1119},
  year={2018},
  publisher={Springer}
}

@inproceedings{cheng2023restoration,
  title={Restoration of atmospheric turbulence-degraded short-exposure image based on convolution neural network},
  author={Cheng, Jiuming and Zhu, Wenyue and Li, Jianyu and Xu, Gang and Chen, Xiaowei and Yao, Cao},
  booktitle={Photonics},
  volume={10},
  number={6},
  pages={666},
  year={2023},
  organization={MDPI}
}

@article{ahmad2022new,
  title={A new generative adversarial network for medical images super resolution},
  author={Ahmad, Waqar and Ali, Hazrat and Shah, Zubair and Azmat, Shoaib},
  journal={Scientific Reports},
  volume={12},
  number={1},
  pages={9533},
  year={2022},
  publisher={Nature Publishing Group UK London}
}

@inproceedings{cohen2018distribution,
  title={Distribution matching losses can hallucinate features in medical image translation},
  author={Cohen, Joseph Paul and Luck, Margaux and Honari, Sina},
  booktitle={International conference on medical image computing and computer-assisted intervention},
  pages={529--536},
  year={2018},
  organization={Springer}
}

@inproceedings{vint2020analysis,
  title={Analysis of deep learning architectures for turbulence mitigation in long-range imagery},
  author={Vint, David and Di Caterina, Gaetano and Soraghan, John and Lamb, Robert and Humphreys, David},
  booktitle={Artificial intelligence and machine learning in defense applications II},
  volume={11543},
  pages={1154303},
  year={2020},
  organization={SPIE}
}

@inproceedings{zhao2023fast,
  title={Fast full-frame video stabilization with iterative optimization},
  author={Zhao, Weiyue and Li, Xin and Peng, Zhan and Luo, Xianrui and Ye, Xinyi and Lu, Hao and Cao, Zhiguo},
  booktitle={Proceedings of the IEEE/CVF International Conference on Computer Vision},
  pages={23534--23544},
  year={2023}
}

@article{lu2025research,
  author    = {Lu, Yijie and Li, Zhengsheng and Qi, Dequan and Zhou, Linhua},
  title     = {Research on Turbulence-Removal Optical Imaging Based on Multi-Scale GAN and Sequential Images},
  journal   = {Optica Open},
  year      = {2025},
  note      = {Preprint},
}

@article{liu2025research,
  title={Research Progress on Atmospheric Turbulence Perception and Correction Based on Adaptive Optics and Deep Learning},
  author={Liu, Qinghui and Di, Yihang and Zhang, Mengmeng and Ren, Zhenbo and Di, Jianglei and Zhao, Jianlin},
  journal={Advanced Photonics Research},
  volume={6},
  number={7},
  pages={2400204},
  year={2025},
  publisher={Wiley Online Library}
}

@article{singh2023robust,
  title={A Robust Basis for Multi-Bit Optical Communication with Vectorial Light},
  author={Singh, Keshaan and Nape, Isaac and Buono, Wagner Tavares and Dudley, Angela and Forbes, Andrew},
  journal={Laser \& Photonics Reviews},
  volume={17},
  number={6},
  pages={2200844},
  year={2023},
  publisher={Wiley Online Library}
}

@article{liu2024atmospheric,
  title={Atmospheric turbulence phase reconstruction via deep learning wavefront sensing},
  author={Liu, Yutao and Zheng, Mingwei and Wang, Xingqi},
  journal={Sensors},
  volume={24},
  number={14},
  pages={4604},
  year={2024},
  publisher={MDPI}
}

@inproceedings{nieuwenhuizen2019deep,
  title={Deep learning for software-based turbulence mitigation in long-range imaging},
  author={Nieuwenhuizen, Robert and Schutte, Klamer},
  booktitle={Artificial intelligence and machine learning in defense applications},
  volume={11169},
  pages={153--162},
  year={2019},
  organization={SPIE}
}

@article{mccann2017convolutional,
  title={Convolutional neural networks for inverse problems in imaging: A review},
  author={McCann, Michael T and Jin, Kyong Hwan and Unser, Michael},
  journal={IEEE Signal Processing Magazine},
  volume={34},
  number={6},
  pages={85--95},
  year={2017},
  publisher={IEEE}
}

@inproceedings{ronneberger2015u,
  title={U-net: Convolutional networks for biomedical image segmentation},
  author={Ronneberger, Olaf and Fischer, Philipp and Brox, Thomas},
  booktitle={International Conference on Medical image computing and computer-assisted intervention},
  pages={234--241},
  year={2015},
  organization={Springer}
}

@inproceedings{liang2021swinir,
  title={Swinir: Image restoration using swin transformer},
  author={Liang, Jingyun and Cao, Jiezhang and Sun, Guolei and Zhang, Kai and Van Gool, Luc and Timofte, Radu},
  booktitle={Proceedings of the IEEE/CVF International Conference on Computer Vision},
  pages={1833--1844},
  year={2021}
}

@inproceedings{lim2017enhanced,
  title={Enhanced deep residual networks for single image super-resolution},
  author={Lim, Bee and Son, Sanghyun and Kim, Heewon and Nah, Seungjun and Mu Lee, Kyoung},
  booktitle={Proceedings of the IEEE conference on computer vision and pattern recognition workshops},
  pages={136--144},
  year={2017}
}

@article{mittal2012making,
  title={Making a “completely blind” image quality analyzer},
  author={Mittal, Anish and Soundararajan, Rajiv and Bovik, Alan C},
  journal={IEEE Signal processing letters},
  volume={20},
  number={3},
  pages={209--212},
  year={2012},
  publisher={IEEE}
}

@article{mittal2012no,
  title={No-reference image quality assessment in the spatial domain},
  author={Mittal, Anish and Moorthy, Anush Krishna and Bovik, Alan Conrad},
  journal={IEEE Transactions on image processing},
  volume={21},
  number={12},
  pages={4695--4708},
  year={2012},
  publisher={IEEE}
}

@article{newbury2024review,
  title={A Review of Differentiable Simulators},
  author={Newbury, Rhys and Collins, Jack and He, Kerry and Pan, Jiahe and Posner, Ingmar and Howard, David and Cosgun, Akansel},
  journal={IEEE Access},
  volume={12},
  year={2024},
  publisher={IEEE}
}

@inproceedings{chakrabarti2016neural,
  title={A neural approach to blind motion deblurring},
  author={Chakrabarti, Ayan},
  booktitle={Proceedings of the European Conference on Computer Vision},
  pages={221--235},
  year={2016},
  volume={9907},
  series={Lecture Notes in Computer Science},
  organization={Springer}
}

@inproceedings{andrychowicz2016learning,
  title={Learning to learn by gradient descent by gradient descent},
  author={Andrychowicz, Marcin and Denil, Misha and Gomez, Sergio and Hoffman, Matthew W and Pfau, David and Schaul, Tom and Shillingford, Brendan and De Freitas, Nando},
  booktitle={Advances in Neural Information Processing Systems},
  volume={29},
  year={2016}
}

@INPROCEEDINGS{10062031,
  author={González, Jorge López and Chapman, Theodore and Chen, Kathryn and Nguyen, Hannah and Chambers, Logan and Mostafa, Seraj A.M. and Wang, Jianwu and Purushotham, Sanjay and Wang, Chenxi and Yue, Jia},
  booktitle={Proceedings of the IEEE/ACM International Conference on Big Data Computing, Applications and Technologies}, 
  title={Atmospheric Gravity Wave Detection Using Transfer Learning Techniques}, 
  year={2022},
  volume={},
  number={},
  pages={128-137},
  publisher={IEEE},
  doi={10.1109/BDCAT56447.2022.00023}
}

@article{UsuiKeisuke2021Qeod,
journal = {Visual Computing for Industry, Biomedicine, and Art},
language = {eng},
number = {1},
pages = {21--29},
publisher = {Springer Science and Business Media LLC},
title = {Quantitative evaluation of deep convolutional neural network-based image denoising for low-dose computed tomography},
volume = {4},
year = {2021},
copyright = {The Author(s) 2021},
author = {Usui, Keisuke and Ogawa, Koichi and Goto, Masami and Sakano, Yasuaki and Kyougoku, Shinsuke and Daida, Hiroyuki},
address = {Singapore},
issn = {2524-4442},
}

@article{jebur2024comprehensive,
  title={A comprehensive review of image denoising in deep learning},
  author={Jebur, Rusul Sabah and Zabil, Mohd Hazli Bin Mohamed and Hammood, Dalal Adulmohsin and Cheng, Lim Kok},
  journal={Multimedia Tools and Applications},
  volume={83},
  number={20},
  pages={58181--58199},
  year={2024},
  publisher={Springer}
}

@article{parvizi2023reinforcement,
  title={Reinforcement learning environment for wavefront sensorless adaptive optics in single-mode fiber coupled optical satellite communications downlinks},
  author={Parvizi, Payam and Zou, Runnan and Bellinger, Colin and Cheriton, Ross and Spinello, Davide},
  journal={Photonics},
  volume={10},
  number={12},
  pages={1371},
  year={2023},
  organization={MDPI}
}

@article{nousiainen2021adaptive,
  title={Adaptive optics control using model-based reinforcement learning},
  author={Nousiainen, Jalo and Rajani, Chang and Kasper, Markus and Helin, Tapio},
  journal={Optics express},
  volume={29},
  number={10},
  pages={15327--15344},
  year={2021},
  publisher={Optical Society of America}
}

\end{document}